\documentclass[12pt,preprint]{aastex}

\shorttitle{Statistics of Polarization and the CF Technique}
\shortauthors{Falceta-Gon\c calves, Lazarian \& Kowal}

\begin{document}

\title{Studies of regular and random magnetic fields in the ISM: statistics of polarization vectors and the Chandrasekhar-Fermi technique}

\author{Diego Falceta-Gon\c calves\altaffilmark{1}}
\affil{N\' ucleo de Astrof\' isica Te\' orica, Universidade Cruzeiro do
Sul - Rua Galv\~ ao Bueno 868, CEP 01506-000 S\~ao Paulo, Brazil}
\affil{Department of Astronomy, University of Wisconsin - 475 N. Charter Street, Madison WI 53706-1582, USA}

\author{Alex Lazarian} 
\affil{Department of Astronomy, University of Wisconsin - 475 N. Charter Street, Madison WI 53706-1582, USA}

\author{and Grzegorz Kowal} 
\affil{Department of Astronomy, University of Wisconsin - 475 N. Charter Street, Madison WI 53706-1582, USA}
\affil{Astronomical Observatory, Jagiellonian University, ul. Orla 171, 30-244 Krak\'ow, Poland}

\altaffiltext{1}{email: diego.goncalves@unicsul.br; diego@astro.wisc.edu}

\begin{abstract}

Polarimetry is extensively used as a tool to trace the interstellar magnetic field projected on the plane of sky. Moreover, it is also possible to estimate the magnetic field intensity from polarimetric maps
based on the Chandrasekhar-Fermi method. In this work, we present results for turbulent, isothermal, 3-D simulations of sub/supersonic and sub/super-Alfvenic cases. With the cubes, assuming perfect grain alignment, we created synthetic polarimetric maps for different orientations of the mean magnetic field with 
respect to the line of sight (LOS). We show that the dispersion of the polarization angle depends on the angle of the mean magnetic field regarding the LOS and on the Alfvenic Mach number. However, 
the second order structure function of the polarization angle follows the relation $SF \propto l^{\alpha}$, $\alpha$ being dependent exclusively on the Alfvenic Mach number. The results show an anti-correlation between the polarization degree and the column density, with exponent $\gamma \sim -0.5$, in agreement with observations, which is explained by the increase in the dispersion of the polarization angle along the LOS within denser regions. However, this effect was observed exclusively on supersonic, but sub-Alfvenic, simulations. For the super-Alfvenic, and the subsonic model, the polarization degree showed to be intependent on the column density. Our major quantitative result is a generalized equation for the CF method, which allowed us to determine the magnetic field strength from the polarization maps with errors $< 20\%$. We also account for the role of observational resolution on the CF method.

\end{abstract}


\section{Introduction}

It is believed that giant molecular clouds in the interstellar medium (ISM) are threaded by large scale magnetic 
fields \citep{sch98, cru99}. 
However, it is still not completely clear what is the role of the magnetic field in the dynamics of the ISM 
and what is its effect on the star formation process. Also, the ratio of the magnetic and turbulent energy in these 
environments is a subject of controversy \citep{padoan02, girart06}. Magnetic fields can influence the injection and evolution of turbulence 
bringing more complexity to this issue (see Lazarian \& Cho [2004] for review). As an example, simulations have shown that 
strongly magnetized turbulent media develop structures with lower density contrasts when compared to pure hydrodynamic 
turbulence \citep{kowal07, kritsuk07}. 

Observationally, different techniques can be used to measure the ISM 
magnetic field and determine its intensity and topology. Zeeman splitting of spectral lines provides a direct and precise 
derivation of the magnetic field component along the line of sight (LOS), mainly for clouds presenting strong spectral lines \citep{heiles05}. 
However, it cannot be applied to the clouds where the line intensities are too weak. 
Typically, when observed, Zeeman measurements of molecular clouds give $B_{\rm LOS} \sim 10^{1-3} \mu$G, and suggest the 
correlation with density $B_{\rm LOS} \propto \rho^{0.5}$, which is consistent to the expected relation for compressions of magnetic fields frozen into plasma. 
Spectral line broadening show that molecular clouds present supersonic, but critically Alfvenic motions \citep{cru99b}. This fact shows 
that the turbulent motions may be excited by MHD modes instead of being purely hydrodynamical.

One of the most readily available methods of studying the perpendicular component of the magnetic field 
is based on the polarization of dust thermal emissions at infrared and submillimetric wavelengths 
\citep{hil00}. The alignment of grains in respect to the magnetic field is a hot research topic 
(see Lazarian [2007] for review). 
Radiative torques (RATs) can promote alignment of irregular dust particles, 
resulting in different intensities for polarized radiation parallel and perpendicular to 
the local magnetic field \citep{dolg76, draine96, draine97, lazho07a}. Grains with long axis aligned perpendicular to the magnetic field induce 
polarization parallel to the magnetic 
field for transmitted star light, and perpendicular to the field lines for the dust emission. 
\cite{cho05} showed that RATs are very 
efficient on the grain alignment process in molecular clouds, even for the very dense regions (up to $A_V<10$). They also showed that the alignment 
efficiency strongly depends on the grain size, being practically perfect for large grains ($a>0.1$ $\mu$m). More detailed studies of the RATs efficiency 
by \cite{lazho07a} confirmed this claim. Therefore, for a range of $A_V$ it is acceptable to assume that the grains are well-aligned.
 
For a given polarization map of an observed region, the mean polarization angle indicates the orientation of the 
large scale magnetic field. On the other hand the polarization dispersion 
gives clues on the value of the turbulent energy. This, as a consequence, can be used to determine the magnetic field component along the 
plane of sky. \cite{chandra53} introduced a method (CF method hereafter) for estimating the ISM magnetic fields 
based on the dispersions of the polarization angle and gas velocity. Simply, it is assumed that the magnetic field 
perturbations are Alfvenic and that the rms velocity is isotropic.

A promising approach to test this method is to create two-dimensional (plane of sky) synthetic maps from numerically simulated cubes.
\cite{ostriker01} performed 3D-MHD simulations, with $256^3$ resolution, in order to obtain polarization maps and study the validity 
of the CF method on the estimation of the magnetic field component along the plane of sky. They showed that the CF method gives 
reasonable results for highly magnetized media, in which the dispersion of the polarization angle is $< 25^{\circ}$. However, 
they did not present any other statistical analysis or predictions that could be useful for the determination of the 
ISM magnetic field from observations. 

\cite{heitsch01} presented a complementary work, with a more detailed analysis regarding the limited observational resolution on the CF method, and presented a modified equation to account for the differences obtained previously. They concluded that lower observational resolution leads to an overestimation of the magnetic field from the CF equation. They also showed a good agreement between the CF technique and the expected magnetic field of their simulations, except for the weak field models. 

Polarization maps from numerical simulations can also be used in the study of the correlation between the polarization 
degree and the total emission intensity (or dust column density). Observationally, the polarization degree in dense 
molecular clouds decreases with the total intensity as $P \propto I^{-\alpha}$, with $\alpha = 0.5 - 1.2$ \citep{goncalves05}.
\cite{padoan01} studied the role of turbulent cells in the 
$P \ versus \ I$ relation using supersonic and super-Alfvenic self-graviting MHD simulations. They found a decrease of polarization degree 
with total dust emission within gravitational cores, in agreement with observations, if grains are assumed to be unaligned for $A_V > 3$. When the 
alignment was assumed to be independent on $A_V$, the anti-correlation was not observed. Recently, \cite{pelkonen07} extended this work and 
refined the calculation of polarization degree introducing the radiative transfer properly. In that work, the decrease in 
the alignment efficiency arises without any {\it ad hoc} assumption. The alignment efficiency decreases as the radiative 
torques become less important in the denser regions. However, it is 
still not clear the role of the magnetic field topology and the presence of multiple cores intercepted by the line of sight 
on the decrease of polarization degree.

In this work we attempt to extend the previously cited studies improving and applying the CF method for different 
situations. For that, we studied both sub and super-Alfvenic models, to study the role of the magnetic field topology in the 
observed polarization maps. We simulate different observational resolutions in the calculations of polarization maps and 
provide combined statistical analysis for both dust absorption and emission maps. We also present statistics based methods to characterize the turbulence and magnetic properties from polarization maps.
We performed numerical simulations of magnetized turbulent plasma with higher resolution, which are described 
in Sec.\ 2. From the data, we computed ``observable" polarization maps, as shown in Sec.\ 3. We then present the 
statistics and spatial distributions of angle and 
polarization degree for different models in Sec.\ 4. In Sec.\ 5 we propose the generalized equation for the CF method and 
compare it with the expected values to study its validity. In Sec.\ 6, we discuss the improved procedures of polarization 
vector statistics, which allow observers to characterize the mean and fluctuating magnetic field of the cloud. We also discuss the applicability of our approach for polarized molecular and atomic lines, and compare the results with previous works. Ou summary is provided in Sec.\ 7.

\section{Numerical Simulations}

The simulations were performed solving the set of ideal MHD equations, in conservative form, as follows:

\begin{equation}
\frac{\partial \rho}{\partial t} + \mathbf{\nabla} \cdot (\rho{\bf v}) = 0,
\end{equation}

\begin{equation}
\frac{\partial \rho {\bf v}}{\partial t} + \mathbf{\nabla} \cdot \left[ \rho{\bf v v} + \left( p+\frac{B^2}{8 \pi} \right) {\bf I} - \frac{1}{4 \pi}{\bf B B} \right] = \rho{\bf f},
\end{equation}

\begin{equation}
\frac{\partial \mathbf{B}}{\partial t} - \mathbf{\nabla \times (v \times B)} = 0,
\end{equation}

\noindent
with $\mathbf{\nabla \cdot B} = 0$, where $\rho$, ${\bf v}$ and $p$ are the plasma density, velocity and pressure, 
respectively, ${\bf B}$ is the magnetic field and ${\bf f}$ represents the external acceleration source, responsible for 
the turbulence injection. For molecular clouds, we may assume that the ratio of dynamical to radiative timescales is very large. Under this assumption, 
the set of equations is closed by an isothermal equation of state $p = c_s^2 \rho$, where 
$c_s$ is the speed of sound. The equations are solved using a second-order-accurate and non-oscillatory 
scheme, with periodic boundaries, as described in \cite{kowal07}.

Initially, we set the intensity of the x-directed magnetic field ${\bf B_{ext}}$ and the gas thermal pressure $p$. This allows us to obtain 
sub-Alfvenic or super-Alfvenic, and subsonic or supersonic models.

The turbulent energy is injected using a random solenoidal function for ${\bf f}$ in Fourier space. This, 
in order to minimize the influence of the forcing in the formation of density structures. We inject energy at scales $k \propto L/l<4$, where $L$ 
is the box size and $l$ is the eddy size of the injection scale. The rms velocity $\delta V$ is kept close to unity, therefore ${\bf v}$ and the Alfv\'en speed $v_A = 
B/\sqrt{4 \pi \rho}$ will be measured in terms of the rms $\delta V$. Also, the time $t$ is measured in terms of the dynamical 
timescale of the largest turbulent eddy ($\sim L/\delta V$). 

We performed four computationally extensive 3D MHD simulations, using high resolution ($512^3$), for different initial conditions, as shown in Table 1. 
We simulated the clouds up to $t_{\rm max} \sim 5$, i.e.\ 5 times longer than the dynamical timescale, 
to ensure a full development of the turbulent cascade. 
We obtained one subsonic and three supersonic models. One of the supersonic models is also super-Alfvenic. Each data cube contains information 
about parameterized density, velocity and magnetic field. As noted from Eqs. (1) and (2), the simulations 
are non self-gravitating and, for this reason, the results are scale-independent.

Regarding the gas distribution in each model we found an increasingly contrast for increasing sonic Mach number, independent on 
the Alfvenic Mach number. This result was also obtained, and studied with more details, in \cite{kowal07}. Subsonic turbulence 
show a gaussian distribution of densities, while the increased number and strnght of shocks in supersonic cases create 
more smaller and denser structures. In these cases, the density contrast may be increased by a factor of 100 - 10000 compared 
to the subsonic case. The magnetic field topology, on the other hand, depends on the Alfvenic Mach number. Sub-Alfvenic models 
show a strong uniformity of the field lines, while the super-Alfvenic case shows a very complex structure. Both effects, the 
density contrast and the magnetic field topology, may play a role on the polarimetric maps, as shown further in the paper.

\section{Polarization maps}

Here we focus on the determination of observable quantities from our synthetic maps. From the density and magnetic field cubes we 
created ``plane of sky" maps for column density and the linear polarization vectors. 
From the velocity field cubes it was possible to obtain the rms velocity, 
which is necessary to test the CF method. 

To create the polarization maps we assumed that the radiation is originated exclusively by thermal emission from 
perfectly aligned grains. The dust abundance is supposed to be linearly proportional to the gas density and, in this case, the total intensity may be assumed to be proportional to the column density. We also assumed that all 
dust particles emit at the same temperature. 

In this work we assume the dust polarization to be completely efficient ($\epsilon = 1$), and perfect grain alignment. Under 
these assumptions, the local angle of alignment ($\psi$) is determined by the local magnetic field projected into the 
plane of sky, and the linear polarization Stokes parameters $Q$ and $U$ are given by: 

\begin{eqnarray}
q = \rho \cos 2\psi \sin^2 i, \nonumber \\
u = \rho \sin 2\psi \sin^2 i,
\end{eqnarray}

\noindent
where $\rho$ is the local density and $i$ is the inclination of the local magnetic field and the line of sight.
We then obtain the integrated $Q$ and $U$, as well as the column density, along the LOS. 
Notice that, for the given equations the total intensity (Stokes $I$) is assumed to be simply proportional 
to the column density. The polarization degree is calculated from 
$P = \sqrt{Q^2+U^2}/I$ and the polarization angle $\phi = atan(U/Q)$. Previous works (e.g. \cite{ostriker01}) 
obtained the polarization maps integrating all cells along the line of sight, in spite of the fact that the local density may be too low to present an observable dust 
component. In reality, dense dust clouds are surrounded by warmer and rarefied regions, in which the dust component is negligible. To simulate this effect, we 
neglect any contribution for cells with density lower than an specific threshold, which depends on the model. The threshold for each model is arbitrarily chosen to 
keep the minimum column density, i.e. intensity, as 0.3 of its maximum.

In Figs. 1 and 2, we show the obtained maps of column density and the polarization vectors for Model 3. We 
used the two extreme orientations of the magnetic field regarding the LOS (0 and $90^{\circ}$). 
Clearly, as shown in Fig.\ 1, the external magnetic field oriented in the x-direction dominates the polarization process. 
Fluctuations on the polarization angle are seen within the condensations, which are dense enough to distort the 
magnetic field lines. In Fig.\ 2, since the magnetic field is oriented along the LOS, the polarization 
is due exclusively by the random field components. The dispersion of the polarization angle is large
and the polarization degree is, in average, lower than obtained in Fig.\ 1. Similar 
results were obtained for Models 1 and 2. Even though presenting different magnetic to gas pressure fraction, 
all the sub-Alfvenic models present similar maps.

In Fig.\ 3, we show the column density and polarization maps of the super-Alfvenic case (Model 4), assuming the magnetic field 
perpendicular to the LOS. Here, the kinetic energy is larger than the magnetic pressure. As a consequence the gas easily tangles the magnetic 
field lines. The angular dispersion is larger and the polarization degree is smaller 
when compared to the sub-Alfvenic case. For the super-Alfvenic case, the orientation of the magnetic field 
regarding the LOS is irrelevant to the polarization maps.

The histograms of polarization angles are shown in Fig.\ 4. In the upper panel we show the histograms for the 
sub-Alfvenic (Models 1, 2 and 3) and the super-Alfvenic (Model 4) cases, with the mean magnetic field lines perpendicular to the LOS. 
The polarization angles present very similar distributions and almost equal dispersion for the sub-Alfvenic cases. This happens mainly because they do not depend on the density structures, but on the 
magnetic topology. Strongly magnetized turbulence creates more filamentary and smoother density structures (i.e.\ low density contrast) if 
compared to weakly magnetized models and, most importantly the magnetic field lines are not highly perturbed. For Model 4, the distribution is practically homogeneous, which means that the polarization is randomly oriented in the plane of sky. It occurs because the turbulent/kinetic pressure is dominant and the gas is able to easily distort the magnetic field lines.

In the bottom panel of Fig.\ 4 we show the polarization angle histograms obtained for Model 3 
but for different orientations of the magnetic field. The dispersion of the polarization angle is very similar for 
inclination angles $\theta < 60^{\circ}$, and increases for larger inclinations. 
It may be understand if noted that the projected magnetic field $B_{sky} = B_{ext} \cos \theta$ is of order of the random component $\delta B$. 
It shows that the dominant parameter that differ the distributions of $\phi$ is the uniform magnetic field projected in the plane of sky, and not the intensity of the global magnetic field.

Furthermore, the distribution pf polarization angles may be used to determine if a sample of clouds in a given region of the ISM present sub or super-Alfvenic turbulence. Since it is very unlikely to have all clouds with mean magnetic field pointed towards the observer, one non-homogeneous distribution of $\phi$ would reveals a sub-Alfvenic turbulence.

\section{Spectra and structure function of polarization angles}

\subsection{Spectra}

In turbulence studies it is useful to calculate the density and velocity power spectra. It allows a better understanding 
and characterization of the energy cascade process and the correlation between different scales. In Fig.\ 5 we present the power 
spectra of the polarization angle for Model 3 with different inclination angles $\theta$ ({\it upper panel}), Model 4 with different $\theta$ ({\it middle panel}) and 
for the different models ({\it bottom panel}). The spectra were obtained for sizes smaller than $L/4$ to 
eliminate contaminations from the forcing at large scales. As the inclination angle $\theta$ increases, more power is 
found in smaller scales (larger $k$) and the spectrum becomes flatter. 

Interestingly, spectrum slopes could be used for the determination 
 of the magnetic field inclination. However, the same trend is found by increasing the sonic and Alfvenic Mach 
numbers, as seen in the bottom panel. 
The degeneracy between Alfvenic Mach number and the magnetic field inclination makes it impossible to correctly estimate $\beta$ 
(or the mass-to-flux ratio) of a given cloud from polarimetric map spectra unless additional information 
regarding the orientation of the magnetic field is given. 
Possibly, a different statistical analysis should be used to bypass this problem. The study of the decorrelation between different scales may 
show more sensitivity to $\theta$ and $M_{\rm A}$ than spectra, as shown below. 

\subsection{Structure functions}

The second order structure function (SF) of the polarization angles is 
defined as the average of the squared difference between the polarization angle measured at 2 points separated by a distance $l$:

\begin{equation}
{\rm SF}(l) = \left< \left| \phi \left({\bf r}+{\bf l }\right) - \phi \left( {\bf r} \right) \right|^2 \right>.
\end{equation}

The structure functions calculated for the different models are shown in Fig.\ 6. In the upper panel we present the SFs 
obtained for Model 3 with different values of $\theta$. As expected, all curves present a positive slope showing the increase in the difference 
of polarization angle for distant points. However, the small scales part of the SF presents a plateau extending up to $l \sim 4 - 5$ pix. This range corresponds to the dissipation 
region and may also be related to the smallest turbulent cells.

As shown in the polarization maps, we should expect an increase in the values of SF as 
$\theta$ increases because of the increase in the dispersion of polarization angles. From Fig.\ 6 (upper panel) it is noticeable the increasing profiles of 
the SFs. However, surprisingly, the obtained slopes are very similar. For an 
assumed relation $SF \propto l^{\alpha}$ we obtained $\alpha \sim 0.5$, for $3 < l < 20$ pix, independently on $\theta$. 

In the middle panel we show the structure functions for Model 4 with different $\theta$. Here, the SFs are almost equal, as noticed by the spectra. We also obtained a very similar slope for the different values of $\theta$. It shows that the slope is independent on $\theta$.

In the bottom panel we show the SFs calculated for the different 
models with $\theta = 0$. It is noticeable the increase in the SF for higher Mach numbers. However, the slopes are 
notably different. The maximum slope is $\alpha \sim 1.1$, 0.8, 0.5 and 0.3 for Models 1, 2, 3 and 4, respectively. Observationally, 
the molecular cloud M17 shows $\alpha \sim 0.5$ up to $l = 3$pc \citep{dotson96}, which would be in agreement with a cloud 
excited by supersonic and sub(or critically)-Alfvenic turbulence. 

From these results we could possibly indicate that SFs of polarization maps may be used for the determination 
of the magnetic field intensity. Associated to other analysis, as spectra and polarization angle histograms, it would be 
possible to determine also the magnetic field inclination regarding the LOS. However, a more detailed study, 
using more numerical simulations considering a large range of parameters, is needed to support these results. 

\subsection{Structure functions at small separations}

As discussed above, the structure functions of polarization angles present a plateau at small scales.
Possibly, if we had ``pencil beam" observations, its range could reflect the size of the smallest turbulent cells ($l_{\rm cell} = l_0$). Infinite resolution observations would 
measure a non-zero (due to the neighboring eddies discontinuity) and flat SF up to a scale $l \sim l_0$. For $l > l_0$ the SF would present a positive slope. However, could the flat part of the SF also be dependent on the observational resolution, instead of the turbulent structures exclusively?

To study this effect, we calculated the polarization maps considering different observational resolutions. 
From the original $512 \times 512$ polarization maps, which is assumed to be the real cloud, we create the 
``observed" maps considering beam sizes $2 \times 2$, $8 \times 8$ and $32 \times 32$ pix. The polarization 
angle is obtained from the $Q$ and $U$ integrated over the neighboring cells, and the SFs for each resolution were obtained from Eq.\ (5). 
The obtained SFs are shown in Fig.\ 7, for $B_{\rm ext}$ perpendicular (upper panel) and parallel to the LOS 
(bottom panel). 

In both plots, we show that the SF is dependent on the observational resolution. As a common result, lower observational resolution results in higher SF at small scales and lower SF at large scales. This is a result of the averaging of the polarization map in the beam size. In the small boxes we show the logarithm of the structure function, and the slopes for the lowest and highest resolutions, 0.35 and 0.50, respectively. Therefore, the observational resolution may influence the determination of the magnetic field from the slopes of SFs, and this method should be used carefully.

An interesting feature is the extended plateau at small scales. The SF present a plateau up to a limiting scale, and a positive slope at larger scales. Observationally, similar profiles were obtained by \citep{dotson96}, which shows that the current results may be implemented by high resolution observations. The limiting scale is approximately the beam size $l \sim l_{\rm res}$. Therefore, the obtained results show that the smallest turbulent cells are only detectable if the condition $l_{\rm res} \ll l_0$ is satisfied in the observations. 
\cite{laz04} estimated $l_0$ for MHD turbulence, considering the viscous and ion-neutral collision damping, as:

\begin{equation}
l_0 \sim \lambda_{in}^{3/4} \left( \frac{c_s}{v_L} \right)^{3/4} \left( \frac{c_A}{v_L} \right)^{1/4} L^{1/4} f_n^{3/4},
\end{equation} 

\noindent
where $\lambda_{in}$ is the mean-free-path for ion-neutral collisions, $v_L$ is the eddy velocity at the injection scale $L$ and $f_n$ is the fraction of neutral atoms. Considering $\lambda_{in} \sim 5 \times 10^{14}$cm, $c_s/v_L \sim 0.1$, $c_A/v_L \sim 1$, $f_n \sim 1$ and $L \sim 100$pc, we obtain $l_0 \sim 10^{-3}$pc. This represents $\sim 0.5$ arcsec for the Orion Molecular Cloud. Considering the instruments available, the required resolution is a little larger than obtained using SHARP and SOFIA ($\sim 2-10$ arcsec), but could be reached by the sub-millimeter array (SMA) ($\sim 0.4$ arcsec). Observations at high resolutions could then also help us to better understand the process occurring at scales smaller than $l_0$. However, the outcome of observations is yet uncertain since \cite{laz04} showed that the magnetic field structures may be complex even below this scale. 

\section{Polarization and CF technique}

\subsection{Polarization degree and column density correlation}

Another interesting analysis is related to the correlation between the column density and the polarization degree. 
As noticeable from Figs. 1, 2 and 3, the polarization degree is smaller within high column density regions 
for all models. This result is supported by observations, and was detected for several objects 
\citep{matthews02, lai03, wolf03}. 
Typically, the polarization degree follows the relation $P \propto I^{-\gamma}$, where $\gamma \sim 0.5 - 1.2$ 
\citep{goncalves05} and $I$ is the total intensity. A possible explanation could be the change on dust size and geometry at denser regions. In this case they would 
be less effectively aligned in respect to the magnetic field \citep{hil99}. Another possibility could be an increase on the thermal collisions with gas and other dust particles 
in high density clumps \citep{laz97}. However, 
our numerical simulations does not take into account such processes and, therefore, those could not 
be influencing our results. 

In Fig.\ 8 we show the correlation between the polarization degree and the column density for Model 3, considering orientations of magnetic field 
regarding the LOS ({\it upper panel}). Also, we show the correlations for the different 
models, assuming the magnetic field at the plane of the sky ({\it bottom panel}). For all angles, the polarization of 
high column densities tend to decrease to the minimum value ($\sim 20\% P_{\rm max}$), which is the value obtained for 
the case of purely random magnetic component ($\theta = 90^{\circ}$). This minimum polarization degree should be zero for 
homogeneous density and random magnetic field. In inhomogeneous media it depends on the number of dense structures intercepted by the line of 
sight. The major contribution for the polarized emission comes from dense clumps, which are few along the LOS. This poor statistics 
results in a non-zero polarization degree. 
For the super-Alfvenic case, the contrast in density is larger as well as the number of high density structures. In this case, the 
polarization degree is smaller, as seen in Fig.\ 8. We obtain, as best fit for the plots, a correlation exponent $\gamma = 0.5$. \cite{cho05}, studying the radiative torque efficiency in the grain alignment process, found larger values for $\gamma$. If grain alignment is implemented properly, the value of $\gamma$ should increase (see Cho \& Lazarian [2005]), but it is out of the scope of this work.
From the bottom panel, it is noticeable that for sub-sonic turbulence the polarization 
degree is large even for the higher column densities. It occurs because in the sub-sonic models the contrast in density is small and the simulated domain is more homogeneous. 
Also, the number of dense clumps, which are able to tangle the field lines, is reduced in the sub-sonic case. On the contrary, for the super-Alfvenic case we obtain a correlation similar 
to $\theta = 90^{\circ}$, i.e.\ purely random magnetic field components. 

These results indicate that 
the decrease in the polarization degree observed in molecular clouds may be partially due to an increase in the random to 
uniform ratio of the magnetic field components. The higher density flows are able to easily tangle 
the magnetic field lines. On the other hand, in the low density streams outside the clumps the magnetic field tends to be 
more uniform and the polarization degree higher. 

\subsection{The CF technique}

\cite{chandra53} proposed a method for estimating the ISM magnetic fields based on the dispersion of polarization 
angles and the rms velocity. Basically, assuming that the magnetic field perturbations are Alfvenic, i.e.\ 
$\delta v \propto \delta B \sqrt\rho$, and that the rms velocity is isotropic we have:

\begin{equation}
\frac{1}{2} \rho \delta V_{\rm LOS}^2 \sim \frac{1}{8 \pi} \delta B^2,
\end{equation}

\noindent
where $\delta V_{LOS}$ is the observational rms velocity along the LOS. Using the small angle approximation 
$\delta \phi \sim \delta B/B_u$, it reduces to:

\begin{equation}
B_u = \xi \sqrt{4 \pi \rho} \ \frac{\delta V_{\rm LOS}}{\delta \phi},
\end{equation}

\noindent
where $\phi$ is measured in radians and $\xi$ is a correction factor ($\sim 0.5$) \citep{zweibel90, myers91}, 
which depends on medium inhomogeneities, 
anisotropies on velocity perturbations, observational resolution and differential averaging along the LOS. 

\cite{ostriker01} noticed from their numerical simulations that the CF method (Eq.\ [7]) was a good approximation 
for the cases where 
$\delta \phi < 25^{\circ}$, i.e.\ when the uniform component of the magnetic field is much larger than the random 
components. This conclusion is expected from Eq.\ (7) since it is applicable 
only for small values of $\delta \phi$, due to the angular approximation.

If one wants to expand the applicability of the CF method for cases where the random component of the magnetic 
field is comparable to the uniform component, or for larger inclination angles, it is necessary to take into account 
two corrections in Eq.\ (7).

Firstly, we must introduce the total magnetic field projected in the plane of sky $B_{\rm sky} 
\sim B_{\rm sky}^{\rm ext} + \delta B$, where $B_{\rm sky}^{\rm ext}$ represents the mean field component projected 
on the plane of sky. We assume here, for the sake of simplicity, that $\delta B$ is isotropic\footnote{This assumption 
is not exact since the magnetic field fluctuations also show anisotropic structures regarding the mean magnetic field. 
Moreover, it was shown that the anisotropy is scale-independent \citep{lp00, esquivel05}.}. \cite{heitsch01} substituted 
$\delta \phi$ in the CF equation by $\delta(\tan\phi)$, where $\tan\phi$ was calculated locally, to provide a correction 
for the small agle approximation. However, they showed that this case lead to an underestimation (by a factor of 100) 
of the magnetic field in super-alfvenic cases. It occurred because, as $|\phi| \rightarrow \pi/2$ it gives 
$B_{\rm CF}^{\rm mod} \rightarrow 0$. To avoid this, they introduced a correction, which was the geometric average 
of the standard $B_{\rm CF}$ and the modified value $B_{\rm CF}^{\rm mod}$. Here, we implement the correction of the 
small angle approximation in a simpler way. We assume that the $\delta B/B$ is a global relation and, in this case, 
we may firstly obtain the dispersion of $\phi$ and then calculate its tangent. Substituting $\delta \phi$ in Eq.\ (7) by $\tan(\delta \phi) \sim \delta B/B_{\rm sky}$, we obtain the modified CF equation: 

\begin{equation}
B_{\rm sky}^{\rm ext} + \delta B \simeq \sqrt{4 \pi \rho} \ \frac{\delta V_{\rm los}}{\tan \left(\delta \phi \right)},
\end{equation} 

\noindent
which is a generalized form of Eq.\ (7). As an example, if polarization maps give
$\delta \phi \rightarrow \pi /4$, Eq.\ (8) gives $B \rightarrow \delta B$ and $B_{\rm u} 
\rightarrow 0$. This is expected for $\theta \rightarrow 90^{\circ}$ or $M_A \gg 1$.

\subsection{Effects of finite resolution}

Here we assume the obtained cubes as the real clouds subject to observational studies. In the previous sections we presented the expected results considering 
infinite observational resolution. However, observational data analysis may be biased by the limited instrumental resolution. Therefore, we must understand 
its role on the statistical analysis of the measured parameters.  

We applied Eq.\ (8) to our simulated clouds, taking into account the effects of finite resolution. Here, 
we intended to determine the role of the resolution on the determination of the magnetic 
field strength from the CF method. We calculated the 
average of the density weighted rms velocity along the LOS ($\delta V_{\rm los}$) and the dispersion of the polarization 
angle ($\delta \phi$) within regions of $R \times R$ pixels. To simulate a 
realistic cloud we chose the mean magnetic field intensities given in Table 2.

In Fig.\ 9 we show the averaged values of the obtained magnetic field for different map resolutions 
($255^2$, $31^2$ and $7^2$ pixels) for Model 3 with different inclinations of the magnetic field.   
For all inclinations, coarser resolution calculations from the CF method tend to 
overestimate the magnetic field intensity. Finer resolutions result in the convergence to the actual 
values $B_{\rm sky}$. 

This trend is seen for different 
inclinations and models. The following equation seems to best fit this behavior:

\begin{equation}
B_{\rm CF} = B_{\rm CF}^0 \left( 1+\frac{C}{R^{0.5}}\right),
\end{equation}

\noindent
where $R$ represents the observational resolution (total number of pixels), $C$ and $B_{\rm CF}^0$ are constants obtained from the best fitting. $B_{\rm CF}^0$ represents the value of 
$B_{\rm CF}$ for infinite resolution observations, i.e.\ the best magnetic field estimation from the CF method. Eq.\ (10) is shown as the dotted lines in Fig.\ 9.

In Fig.\ 10 we show the magnetic field obtained from Eq.\ (8) for the different models with $\theta = 0$. The dotted 
lines represent the best fitting using Eq.\ (10). 

The fit parameters, as well as the 
expected values of the magnetic field from the simulations for all models, are shown in Table 2. Here, the 
magnetic fields are given in units of the mean field $B_{\rm ext}$. Since the simulations are scale independent, 
one could choose values of $B_{\rm ext}$ to represent a real cloud, in accordance with the parameters of Table 1. As an 
example, assuming a cloud with $n_{\rm H} = 10^3$cm$^{-3}$ and $T=10$K, and $\beta = 0.01$ (Model 3), we get 
$B \sim 50\mu$G. Choosing differently the density, temperature or the model given by the simulations, i.e.\ $\beta$,
 we obtain a different mean magnetic field. 
The obtained parameter $C$ is very similar for the different inclinations, but are different depending 
on the model, mainly because it is related to the scale on which the dispersion of the polarization angle changes. Since 
$C$ seems to depend on the model and not on the inclination it could also be used by observers to infer the physical 
properties of clouds from polarization maps. 

It is shown that
$B_{\rm CF}^0$ decreases as $\theta$ increases but do not reach zero as would be expected from $B_{\rm sky}^{\rm ext}$. 
However, if we compare the magnetic field strength obtained from the CF method with the total magnetic field (last 
column of Table 2) as proposed in Eq.\ (8), the convergence between the estimative and the actual values is much better. The error
($\epsilon = [log(B_{\rm CF}^0)-log(B)]/log(B_{\rm CF}^0)$) using this method is $<20\%$ considering all cases, 
validating the CF method under the assumptions used for Eq.\ (8).

As a practical use, observers could obtain polarimetric maps of a given region of the sky for different observational 
resolutions (e.g.\ changing the resolution via 
spatial averaging). Using the CF technique for each resolution and, then 
apply Eq.\ (9) to determine the asymptotic value of the magnetic field projected into the plane of sky $B_{\rm CF}^0$. 

\subsection{Polarization of stellar radiation}

Dense cloud envelopes and diffuse clouds typically present very weak or no far-IR and sub-mm dust emission. 
In these cases, it is very difficult to obtain polarimetric maps from dust emission and other methods are necessary. 
Some of these clouds are known to intercept rich clusters of stars. IR, optical and ultraviolet (UV) emissions from 
these stars suffer extinction by the dust component of the intercepting clouds and, as a consequence, the detected 
stellar radiation may be polarized. 

Absorption polarization maps are considered ``infinite resolution" measurements of polarization vectors and 
can be very useful on the study of the magnetic field of the ISM. However, since it depends on the stellar 
background, detections are rare and the polarization maps are sparse. 

To test if the current absorption polarization maps are statistically relevant, 
as well as the applicability of the CF technique for these type of observation, we simulated the polarization 
of background stars in our cubes. Assuming our cubes to be 1 arcmin$^2$ of the sky, and using the density 
estimative of $10^3$/arcmin$^2$ for our 
Galaxy \citep{garwood87}, we recalculated the polarization maps of $10^3$ randomly positioned stars. 
Each star is assumed to originate an unpolarized total intensity $S$, which is absorbed by the dust component intercepting the line of sight. 
At each cell, we compute the absorbed intensities $\delta I_{\rm x}$ and $\delta I_{\rm y}$, which depend on the local magnetic 
field orientation. Again, as in the calculation of the emission maps, we assume maximum efficiency in the polarization by 
the dust. After integration, the total absorptions $I_{\rm x}$ and $I_{\rm y}$ are used to obtain $Q$ and $U$.
The obtained polarization angle is then rotated by $90\degr$ (opposite polarization) in order to be compared with the emission 
polarimetric maps.

In Fig.\ 11 we illustrate the obtained results. We exemplify the obtained maps with a zoomed clumpy 
region (100 x 100 pixels) of Model 3 with $\theta = 90\degr$ (upper panel). Here, it is shown that 
just a few stars ($<50$) are detected. 

To test the predictions of the magnetic field proposed in this work, we calculated the dispersion and the structure 
function of the polarization angles. In the bottom panel of Fig.\ 11 we show the histogram of $\phi$, and its structure 
function (squares), for Model 3 with $\theta = 0\degr$. We compared it with the dust emission SF (solid line). 
The structure functions seem very similar but, due to the small number of stars, the dispersion in the SF for absorption 
is large. In this case, the SFs from absorption maps may present too large uncertainties, which make difficult 
the analysis of the magnetic field from SF slopes.

An alternative would be the use of the improved CF technique presented in Sec.\ 5. 
We applied the CF method to the absorption polarization maps of Model 3, with $\theta = 0\degr$, which resulted in 
$B_{CF}^{\rm abs} = 550 \pm 126$ $\mu$G. This value is comparable to the result obtained from the emission 
polarimetric maps ($B_{CF}^{\rm em} = 464 \pm 45$ $\mu$G). Even with a small number of stars, the obtained result 
is similar to the finest resolution case of polarized emission. It occurs because the stars act as single pixel measurements 
and there is no averaging of $\phi$ within the observational beam size. 
The down-side of this technique is its higher noise. 

Currently there are few observed polarization maps available from extinction of background stars. Fortunately, some projects are being implemented and promise to bring us complete sets of polarization maps of background stars, which would be compared to the presented results, like the Galactic Plane Infrared Polarization Survey \citep{shiode06}.
However, we believe that joint analysis of both, emission and absorption, polarization maps can provide a more complete understanding of the magnetic field in dense and diffuse regions of the ISM.

\section{Discussions}

Emission and extinction polarimetric measurements provide an unique technique for the study of the magnetic field, 
projected into the plane of sky, in molecular clouds. Synthetic extinction maps depend on additional assumptions about the 
stellar population and may be more explored in a future work. 
In this work we focused on providing synthetic emission 
polarimetric maps, as well as different statistical analysis that could be used in the future by observers to infer 
the physical properties of the studied region. The physical interpretations of our results, as well as the comparisons with 
previous theoretical works, are given as follows.

\subsection{Our models}

In this work we presented four different models: (1) $\beta = 1.0$, $M_{\rm S}=0.7$ and $M_{\rm A}=0.7$, (2) $\beta = 0.1$, 
$M_{\rm S}=2.0$ and $M_{\rm A}=0.7$, (3) $\beta = 0.01$, $M_{\rm S}=7.0$ and $M_{\rm A}=0.7$ and (4) 
$\beta = 0.1$, $M_{\rm S}=7.0$ and $M_{\rm A}=2.0$. Similar studies provided by \cite{ostriker01} characterized 
different models by their pressure ratio. However, our results show completely different polarization maps for 
the two coincident $\beta$-value models. This because the super-Alfvenic flows tend to tangle the magnetic field lines, 
what is not seen in the sub-Alfvenic models (Model 2), even with similar pressure ratio.

The super-Alfvenic case shows a randomly distributed column density maps, with high constrast between the denser and rarefied regions. 
On the other hand, sub-Alfvenic cases are more filamentary, with contrasts increasing with the sonic Mach number. 
This general picture is independent on the angle between the external mean magnetic field and the LOS $\theta$. 
However, for $B_{ext}$ nearly parallel to the LOS, the observed polarization 
will mostly be due to the random fluctuation component $\delta B$. This effect is noticeable comparing Figs. 1 and 2.

We see that for sub-Alfvenic turbulence the large scale density enhancements are mostly parallel to the mean magnetic fields, 
with exception to the very dense cores, which can easily change the orientation of the magnetic field. As a consequence, 
polarization maps will present dense structures mostly aligned with the mean magnetic field. This effect also play a role on 
the generation of the polarization maps. Since we integrate the polarization vectors along the LOS, the low density cells 
will systematically increase the homogeneous contribution, as well as the resulting polarization degree. To avoid this effect, 
we disregarded the contribution from low density cells using a threshold, which depends on the model used.
For the models where the magnetic field is oriented parallel to the LOS, the polarization maps will show polarization vectors randomly 
oriented in respect to the density structures. It reveals the degeneracy on the polarimetric maps between the super-Alfvenic models with those with 
$B$ nearly parallel to the LOS.

\subsection{Polarization degree versus emission intensity}

The polarization maps showed that the polarization degree is anti-correlated 
to the column density, in exception to the subsonic case. This result is in agreement 
with the observations, which revealed ``polarization holes" associated to the 
dense cores for most of the regions observed. 

Observationally, \cite{wolf03} showed that the polarization maps of molecular clouds 
follow the relation $P \propto I^{-\gamma}$, with $\gamma \sim 0.5 - 1.2$. The same trend is observed from polarized extinction of background stars \citep{arce98}. It was proposed that this behavior occurs 
due to changes on dust properties inside denser cores, or even by an increase in thermal pressure, causing depolarization. 
\cite{cho05} studied the role of the radiative torques on the grain alignment at dense cores and obtained $\gamma \sim 0.5 - 1.5$, 
depending on the dust size distribution. 

\cite{padoan01} obtained a similar 
behavior from their numerical simulations of protostellar cores, though for only three dense cores of 
one single simulation. They assumed a cut-off on grain alignment efficiency for $A_V > 3$mag. For their model, 
if the alignment efficiency is independent on $A_V$, the polarization degree was shown to be independent on the 
column density. \cite{pelkonen07} extended this work, improving the 
radiative transfer. They naturally obtained a decrease in the grain alignment at denser regions, explaining the lower 
degree of polarization.

In our models, we assumed perfect grain alignment (independent on $A_V$). Therefore, 
the depolarization is exclusively due to the dispersion increase of the polarization angles in denser regions.
We obtained $\gamma \sim 0.5$ for Models (2) and (3), but no correlation (i.e.\ $\gamma \sim 0$) 
was found for Models (1) and (4). For Model (1), even the densest cores are unable to tangle the magnetic field lines 
and the polarization degree is homogeneously large. For Model (4), we have the opposite situation. The super-Alfvenic 
turbulence causes a strong dispersion of the magnetic field even at the less dense regions. For this case, the polarization 
degree is low everywhere. For Models (2) and (3), the turbulence is unable to destroy the magnetic field structure, but is 
able to create the dense cores by shocks. The cores are dense enough to drag the magnetic field lines and to increase the 
local $M_A$.

Possibly, our results differ from the obtained by \cite{padoan01} because of: i - numerical resolution, 
ii - $M_A$, and iii - self-gravity. We used $512^3$ simulations (instead of a $128^3$) and, as a result, our magnetic 
field structure is less homogeneous and the density constrast is higher. The larger complexities present in our cubes 
increase the effect described in the previous paragraph. \cite{padoan01} and \cite{pelkonen07} used a single, super-Alfvenic, 
model. We showed that the polarization maps for $M_A$ present a flat $P \times I$ correlation. Finally, self-gravity causes the 
colapse of the denser regions compressing the magnetic field within these cores. As a consequence, if no strong diffusion 
takes place, the polarization degree tends to grow.
As a future work, we plan to study properly the depolarization at dense cores considering the grain alignment process and 
self-gravity.

\subsection{Statistics of polarization angles}

We found that the distributions of polarization angles of sub-Alfvenic models are similar, even for 
different magnetic to gas pressure ratios. However, the dispersion of angles increases with $M_{\rm A}$ and with the inclination 
of the external magnetic field regarding the line of sight ($\theta$). Actually, we noticed that the critical parameter is 
the Alfvenic Mach number considering the magnetic field component projected into the plane of sky (i.e. $M_{\rm A}^{\rm sky} = 
\delta v \sqrt{4 \pi \rho}/B_{\rm sky}$). We can compare these results with \cite{padoan01} and \cite{ostriker01}. The first 
used one single super-Alfvenic model, and obtained irregular (flat) distributions of polarization angle. The latest analysed 
supersonic models for $\beta = 0.01, 0.1$ and 1.0. They obtained clearly gaussian distributions for $\beta = 0.01$, with 
increasing dispersion for larger inclinations. Also, they obtained flatter distributions as $\beta$ increases (i.e. as 
the Alfvenic Mach number increases). These are all in agreement with our results.

On the other hand, the power spectra analysis was showed to depend on the sonic Mach number. 
The spectra of the polarization angles show an increase in the power of small scales for increasing $M_{\rm S}$. 
This occurs due to the amplification on the perturbations of the smallest scales for stronger turbulence. 
However, the same behavior is seen varying the inclination of the 
mean magnetic field. In this sense, there is a degeneracy between the Alfvenic Mach number and the orientation of 
the magnetic field.
In this sense, structure functions of the polarization angle showed to be useful to avoid this 
degeneracy. 

The sub-Alfvenic models presented SFs with slope $\alpha \sim 0.5$ 
($SF \propto l^{\alpha}$), independent on the magnetic field 
orientation. On the other hand, the super-Alfvenic model presented flatter SFs ($\alpha \sim 0.3$). Therefore, we 
could conjecture that it could be possible to obtain the 
mean magnetic field intensity, independently on its orientation regarding the LOS, from the SF slopes. 
Needless to say that more models are needed to 
confirm this possibility. 

Besides, SFs can potentially be used to study the turbulence eddies. Its flat profile at small scales may provide informations about 
the amplitude and size of the smallest turbulence cells. However, we showed that the results are affected by 
the observational resolution.

\subsection{Improved CF technique}

The CF method has been proven an useful tool for the determination of the magnetic field in the ISM. 
We studied its validity using the obtained polarization maps from our models. 
\cite{ostriker01} proposed that the CF method would only be applicable in restricted 
cases, in which the dispersion of the polarization angle is small ($< 25^{\circ}$). It is consistent with the original 
approximations involving the derivation of the CF equation. We derived a generalized formula for the CF method, 
based on the same assumptions of the original work \citep{chandra53}, but that accounts for larger dispersion models. 
Basically, we assume that the perturbations are Alfvenic and that we have an isotropic distribution of 
$\delta V$. We, for the first time, successfully applied this equation to the super-Alfvenic and large inclination models. 

We also studied dependency of the CF method and the observational resolution. As also shown by \cite{ostriker01} and 
\cite{heitsch01}, the CF method overestimates the magnetic field for coarser resolutions. Therefore, we propose 
a general equation to fit the observational data considering maps with different resolution. The asymptotic 
value of the given procedure provides the ``infinite resolution" measurement from the CF method and is consistent 
with the expected values from the simulations. 

As stated before, a possible limitation in the presented model is the absence of self-gravity effects. 
At the denser regions, the magnetic field configurations may possibly be different as the cloud collapses and drags 
the field lines. This process is responsible for the hour-glass structures observed in several gravitationally unstable 
clouds (e.g. Vall\'ee \& Fiege 2007). As a consequence, self-gravity increases the magnetic field locally 
and reduces the dispersion of the polarization angles within the dense clumps. To test the stability of the dense clumps 
in our simulations we may estimate the Jeans length ($\lambda_J = c_s \sqrt{\pi/G\rho}$). For the parameters chosen in 
Section 5.3, the denser structures, considering all models, are characterized 
by $l_{\rm core} \sim 0.1 - 0.5$pc, $n_{\rm core} \sim 10^{5-6}$cm$^{-3}$ and $T=10 - 100$K, resulting in 
$\lambda_J \sim 0.1 - 1$pc. Therefore, since $l_{\rm core} \sim \lambda_J$, the denser structures may be unstable, 
at least for the given parameters. On the other hand, self-gravity plays a role at small regions and may not 
be statistically important for the previous results (except for the $P \times I$ correlation) as we studied regions 
much larger than the very dense clumps. 
We plan to study the effects of self-gravity on the obtained results in a further 
work.

\subsection{Sub-Alfvenic versus super-Alfvenic turbulence}

The ratio of thermal gas to magnetic pressures ($\beta$) is typically used as the dominant parameter on 
the characterization of the degree of magnetization of a cloud. In this sense, systems with similar 
$\beta$ values should present similar distribution of structures and dynamics. However, we showed that 
the sonic ($M_{\rm S}$) and Alfvenic ($M_{\rm A}$) Mach numbers, which quantify the ratio of the kinetic to the thermal and 
magnetic pressures, respectively, divide the models in different regimes. For the case of polarization 
vectors, our simulations showed that $M_{\rm A}$ is decisive. 

For clouds with $M_{\rm A} < 1$, the gas motions excited by turbulence are confined by the magnetic field and are 
not able to change its configuration. Actually, the perturbations in the magnetic field occur, but are 
small compared to the mean field ($\delta B \ll B_{\rm m}$). In this case, the polarization vectors are uniform, 
as shown in Fig.\ 4. For $M_{\rm A} > 1$, the magnetic pressure is small compared to the kinetic energy of the turbulent 
gas and the mean magnetic field can be easily distorted. As a consequence, the polarization maps would show large 
dispersion of $\phi$. 

Obviously, a large dispersion of $\phi$ can also be related to a projection effect. If the mean magnetic field 
is projected along the line of sight ($\theta \sim 90\degr$), only the small perturbations $\delta B$ will be seen 
as the polarization vectors. 
However, considering a large number of clouds, there is a very low probability for all to present $\theta \sim 90\degr$. 
In this case, if observations systematically show very large dispersions of $\phi$, it means that the turbulence 
in the ISM may be typically super-Alfvenic. Otherwise, the ISM then presents sub(quasi)-Alfvenic turbulence. 
Observations of a given cloud could then be compared to our Fig.\ 4 to determine under which regime the turbulence 
is operating. It is particularly interesting since the ratio of magnetic to turbulent energy in the ISM is still subject 
of controversy \citep{padoan02, girart06}.

\subsection{Procedure for observational data analysis}

The number of simulations presented in this work, as well as the numerical resolution, must be increased in future works. 
In any case, from the models we have, we intend to provide observers with a straightforward procedure to characterize 
the magnetic field and turbulence properties of molecular clouds.

Firstly, from the polarization maps of a given region one should obtain the second order structure function of the 
polarization angle. From the SF, it is possible to characterize the turbulent cascade and the magnetic field. The 
extension of the flat profile at small scales give the turbulence cut-off scale. On the other extreme, the flat profile 
at large scales indicate the energy injection lengths. From the maximum slope of $SF \propto l^{\alpha}$, it is possible 
to determine the averaged Alfvenic Mach number and, as a consequence, the magnetic field intensity. 

Another method to obtain the magnetic field intensity is based on the CF technique. From the observed velocity dispersion 
it is possible to estimate the amplitude of the random component of the magnetic field from Eq.\ (7). The total magnetic 
field is then obtained from Eq.\ (9), using the dispersion of the polarization angle. To avoid the dependence on the 
observational resolution, it is suggested to evaluate the dispersion of the polarization angle for different resolutions 
(which may be simulated by averaging neighboring vectors of the polarization maps) and determine the asymptotic total 
magnetic field from Eq.\ (10). Finally, subtracting the total field by the random component, it is possible to 
determine the mean magnetic field projected in the plane of sky. 

Also, combining the mean magnetic field obtained from both methods, it is possible to estimate the angle between 
the mean magnetic field and the LOS. 

\subsection{Grain alignment}

Although it was not considered in the present calculations, a correct treatment of grain alignment is needed for a full understanding of the polarization in molecular clouds. For instance, we showed that the polarization degree is anti-correlated with the column density with slope $\gamma \sim -0.5$, while observations sometimes give $\gamma < -1.0$ \citep{goncalves05}. This difference is related with the alignment efficiency at different regions of the cloud \citep{cho05}.

The theory of grain alignment has developed fastly during the past decade (see Lazarian [2007]). It is currently believed that radiative torques play a major role on the alignment process and it strongly depends on $A_V$ (see Lazarian \& Hoang [2007]). With increasing extinction ($A_V > 2$), the radiative torques are less effective and only large grains are aligned. All in all, both observations \citep{arce98, whittet08} and theory \citep{hoang08}, suggest that there is a range of $A_V$ for which our assumptions are correct. It might happen that subsonic mechanical alignment of irregular grains, introduced in \cite{lazho07b}, extends the range of $A_V$ over which grains are aligned when compared to the estimates based on radiative torques only.

The observed band is also selective regarding the dust sizes and different bands reveal the polarization of different dust components. All these effects will be included in a future work, and a more realistic study of the polarization intensity distribution will be obtained.

\subsection{Polarization from molecular and atomic lines}

In the present work we focused on calculating synthetic polarization maps of FIR emission from dust particles, 
which were assumed to be perfectly aligned with the magnetic field. Unfortunately, due to inefficient grain alignment 
at the dense cloud cores, the dust polarization degree may decrease and different methods have to be used. 

The polarization of molecular lines have been shown 
to be an additional tool for the study of the magnetic fields in the ISM \citep{girart99,greaves02, girart04,cortes05}. 
Molecules are present in the dense and cold cores of the clouds and may be detected by thermal line
emissions. Polarimetric maps of molecular emission can be used on the study of regions with $A_V > 10$, 
and can be directly associated with the Zeeman measurements. Based on the Goldreich-Kylafis effect \citep{gold81,gold82}, 
the molecular sublevel populations will present imbalances due to the magnetic field, generating polarized rotational 
transitions. However, the survival of molecules depend on restrict conditions, as for $A_V > 10$ 
molecules may be frozen into dust particles. Another difficulty regarding this method 
is the fact that the GK effect generates polarization either parallel or perpendicular to the magnetic field lines, 
making the polarization maps.

On the other hand, polarized scattering and absorption from atoms and ions provide information about the magnetic field 
in warm and rarefied regions, like the diffuse ISM and the intergalactic medium. 

Polarization arising from aligned atoms and ions is a new method \citep{yan07a}. Unlike molecular lines that live in the excited state long enough to be imprinted by the magnetic field, the atomic exited states are short lived and tend to decay in timescales shorter than the Larmor precession of the atom. However, species with fine and/or hyperfine structure of ground or metastable states can be aligned. This fact opens new horizons for polarimetric studies of magnetic fields \citep{yan06, yan07a, yan07b}.

It is useful to comment here that the results shown in this work are also valid for the observed polarization maps of molecular and atomic emission lines. This because the assumptions made for the calculations disregard any special consideration about the emitting species, which could be atoms, molecules or dust particles.

Also, it is worth mentioning that these techniques are either a substitute, for regions where no FIR dust emission is detectable, or complementary to the dust polarized emission but at different wavelengths (e.g.\ optical and UV radiation). Depending on the $A_V$ range considered, polarization of atoms and molecules may complement the dust emission and absorption maps, which are usually much more detailed.

\subsection{Comparison with previous works}

In this work we studied of polarization maps and its applicability on the determination of magnetic fields 
in molecular clouds based on numerical simulations. Here we compare the obtained results with the previous 
theoretical works.

\cite{ostriker01} performed numerical simulations, with $256^3$ resolution, considering plasma 
$\beta$ values 0.01, 0.1 and 1.0. Their results showed homogeneous polarization maps 
for $\beta = 0.01$ (strongly magnetized turbulence), and a complex distribution of polarization vector for 
$\beta = 0.1$ and 1.0 (weakly magnetized turbulence). They also obtained an increase in the dispersion of 
polarization angles with the increase of the external magnetic field inclination regarding the line of sight. These 
results are in agreement with our models, except for the fact that two of our models with equal $\beta$ 
presented completely different polarization maps. This because the $\beta$ value does not reveal how the magnetic field lines 
respond to the turbulence. The Alfvenic and sonic 
Mach numbers reflect how strong is the turbulent pressure compared to the magnetic and thermal pressures, respectively. 

They obtained a higher polarization degree for $\beta$ = 0.01, obviously because of the magnetic field intensity, 
but larger values of $P$ for larger column densities (i.e.\ for larger $I$), in disagreement with observations. We believe 
that this was caused by their method for obtaining the polarization degree. They obtained the integrated Stokes parameters, 
weighted by local density, for all cells along the LOS. We, on the other hand, used a threshold on density to avoid the 
contribution of very rarefied regions (where the magnetic field structure is systematically more uniform). 
\cite{padoan01} focused their work on the polarization of dust emission from dense cores. 
They implemented a more realistic calculation of the polarization degree based on the efficiency of the alignment 
for different values of $A_{\rm V}$. In this case, they were able to obtain a decreasing polarization degree with 
the total intensity related to the grain properties, and not to the statistics of polarization vectors along the line of 
sight. We also believe that the numerical resolution may be playing a role on the polarization degree. 
More refined simulations systematically result in more complex structures for density and magnetic field 
lines. As a consequence, the alignment vectors along the line of sight present larger dispersion resulting in a lower 
polarization degree. 

We found no previous theoretical work presenting an extended statistical anaylsis considering all the 
PDF, Spectra and Structure Function of polarization angle, and therefore no comparison can be made. 

\cite{ostriker01} and \cite{padoan01} also tested the CF technique using their simulations, considering Eq.\ (8). 
They obtained good agreement, with a correction factor of $\sim 0.5$, between the calculated estimations and the 
expected values only for the models with $\delta \phi < 25\degr$. 
\cite{heitsch01} presented an extended analysis using a larger number of models, with different physical parameters 
and numerical resolutions (including 1 model with $512^3$ resolution). They also studied the effects of observational 
resolution on the obtained maps. They concluded that coarser resolutions result in more uniform polarization vectors. 
As a consequence, the CF method overestimates the magnetic field intensity. This is in full agreement with our results. 
They also tested the reliability of the CF technique in weakly magnetized clouds. They proposed the modified equation 
$B_{CF}B_{CF}^{mod}=4\pi \rho [\delta v_{\rm LOS}/\delta(\tan \phi)][1+3\delta(\tan \phi)^2]^{1/2}$, 
\footnote{Here, $B_{CF}$ is the value obtained using the standard CF equation, and $B_{CF}^{mod}$ is the corrected 
value proposed by \cite{heitsch01}.} which gave good results compared with the expected values for their models, with 
discrepancies of a factor $<2$.

We tested their equation to our models 3 and 4 with $\theta = 0$, representing a strong and 
weakly magnetized cloud, respectively. For Model 3, the obtained value is in agreement with that shown in 
Table 2. The ratio between the two measurements is $B_{CF}^{mod}/B_{\rm CF}^0 = 0.9$. For Model 4, the proposed equation 
underestimates the magnetic field, and compared to with our method it gives $B_{CF}^{mod}/B_{\rm CF}^0 \sim 0.3$. 
Despite of the few simulations available for compariron, their method seems to systematically underestimate 
the magnetic field intensity. More tests are needed to determine which method may give the best results.

Furthermore, even though not addressed by \cite{heitsch01}, we studied the dependence of the polarization angles and 
the CF technique with the inclination of the magnetic field regarding the LOS. We showed that there is a degeneracy 
between the results of weakly magnetized clouds and strongly magnetized clouds with high $\theta$. The modified CF formula 
presented in this work gave good results for all cases.

As discussed before, even though not taking into account the self-gravity in our simulations it mostly 
induces changes at small scales, as noted by 
\cite{heitsch01}. As a result, they showed the CF technique to be insensitive to self-gravity.

\section{Summary}

In this work we presented turbulent 3-D high resolution MHD numerical simulations in order to study the 
polarized emission of dust particles in 
molecular clouds. We obtained synthetic dust emission polarization maps 
calculating the Stokes parameters $Q$, $U$ and $I$ 
assuming perfect grain alignment and that the dust optical properties are the same at all cells. 
Under these conditions, we were able to study the polarization angle distributions and the polarization degree for the 
different models and for different inclinations of the magnetic field regarding the LOS. 
As main results, we:

\medskip
- obtained an anti-correlation between the polarization degree and the column density, 
with exponent $\gamma \sim -0.5$, related to random cancelation of polarization vectors integrated along the LOS;

\medskip
- showed that the overall properties of the polarization maps are related to the Alfvenic Mach number and 
not to the magnetic to gas pressure ratio. 

\medskip
- calculated the spectra and structure functions of the polarization angles, and obtained degenerate 
conclusions for the Alfvenic Mach number and the angle between the magnetic field and the LOS;

\medskip
- presented a generalization of the CF method, which showed useful for: i- the determination of the total 
magnetic field projected in the plane of sky, and ii- the separation of the two components $B_{\rm sky}$ and $\delta B$;

\medskip
- studied the effects of different observational resolutions on the CF method. We presented an empirical 
equation to determine the correct magnetic field from different resolution measurements;

\medskip
- studied the effects of different observational resolutions on the structure function of the polarization angle. We 
discuss the applicability of SFs for the determination of turbulent cut-off scales.

\medskip
These results represent important tools for present and future polarimetric FIR observational studies.
In the future it would be necessary to increase the number of simulated models, 
with different physical parameters and with better resolutions, to test these conclusions. It would also be interesting to 
implement the grain alignment processes properly and study their effects on the obtained results.
Using more computational models it will be possible to test the 
proposed method to remove the degeneracy between the Alfvenic Mach number and the angle between the magnetic field 
and the LOS. This would also help us to provide the observers a large number of simulated clouds 
that could be used as benchmarks. 

\acknowledgments

D.F.G., A.L. and G.K. thank the financial support of the NSF (No.\ AST0307869), the Center for Magnetic Self-Organization 
in Astrophysical and Laboratory Plasmas and the Brazilian agencies FAPESP (No.\ 06/57824-1 and 07/50065-0) and CAPES (No.\ 4141067).

\clearpage
\begin{table}
\begin{center}
\caption{Description of the simulations - MHD, 512$^3$ \label{table1}}
\begin{tabular}{cccccc}
\hline\hline
Model & $P$ & $B_{ext}$ & \multicolumn{1}{c}{$M_S$\tablenotemark{a}} & $M_A$\tablenotemark{b} &Description \\
\tableline
1 &1.00 &1.00 &$0.7$ &$0.7$ & subsonic \& sub-Alfvenic\\
2 &0.10 &1.00 &$2.0$ &$0.7$ & supersonic \& sub-Alfvenic\\
3 &0.01 &1.00 &$7.0$ &$0.7$ & supersonic \& sub-Alfvenic\\
4 &0.01 &0.10 &$7.0$ &$2.0$ & supersonic \& super-Alfvenic\\
\hline\hline
\end{tabular}
\tablenotetext{a}{sonic Mach number ($M_S = \left< v/c_S \right>$)}
\tablenotetext{b}{Alfvenic Mach number ($M_A = \left< v/v_A \right>$)}
\end{center}
\end{table}

\clearpage
\begin{table}
\begin{center}
\caption{CF method estimates \label{table2}}
\begin{tabular}{cccccc}
\hline\hline
Model & $\theta (^{\circ})$ & $C$ & $B_{\rm CF}^0/B_{\rm ext}$ & $B_{\rm sky}^{\rm ext}/B_{\rm ext}$ \tablenotemark{a} & $B_{\rm tot}/B_{\rm ext}$ \tablenotemark{b}\\
\tableline
3 & 0 &$20\pm5$ &$1.24 \pm 0.09$ & 1.00 & 1.25\\
3 & 30 &$24\pm5$ &$0.98 \pm 0.08$ & 0.87 & 1.11\\
3 & 45 &$25\pm5$ &$0.78 \pm 0.07$ & 0.71 & 0.96\\
3 & 60 &$33\pm5$ &$0.48 \pm 0.05$ & 0.50 & 0.75\\
3 & 90 &$31\pm5$ &$0.26 \pm 0.03$ & 0.00 & 0.24\\
\hline
1 & 0 &$7\pm5$ &$0.97 \pm 0.08$ & 1.00 & 1.11\\
2 & 0 &$10\pm5$ &$1.07 \pm 0.07$ & 1.00 & 1.16\\
4 & 0 &$34\pm5$ &$1.18 \pm 0.07$ & 1.00 & 1.41\\
\hline\hline
\end{tabular}
\tablenotetext{a}{Mean field adopted for the model, projected into the plane of sky, i.e.\ $B_{\rm sky}^{\rm ext}=B_{\rm ext} \cos \theta$}
\tablenotetext{b}{Total field of the model, projected into the plane of sky, i.e.\ $B_{\rm tot}=B_{\rm sky}^{\rm ext}+\delta B$}
\end{center}
\end{table}

\clearpage
\begin{figure*}
   \centering
   \includegraphics[width=12cm]{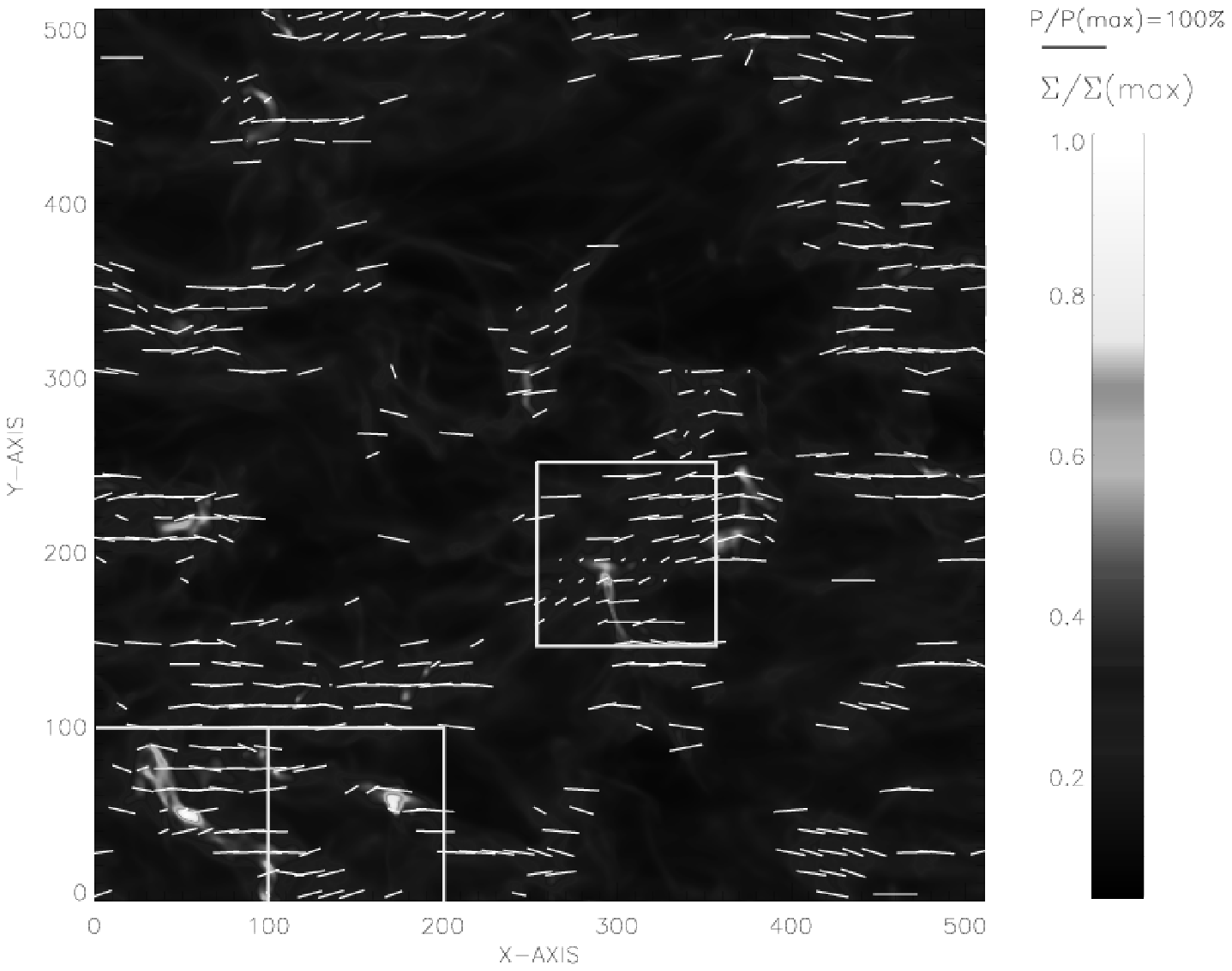}
   \caption{Polarization of emission and column density maps for Model 3 ($M_{\rm S} \sim 7.0$  and $M_{\rm A} \sim 0.7$) with ${\bf B}_{ext}$ perpendicular to the line of sight. The complete map (512x512 pix) ({\it Upper-left}) and the zoomed regions (100x100 pix). The sensitivity in simulated observations is assumed to be 0.3 of the maximum emission. Here, regions where the signal is less than 0.3 do not show polarization vectors, and $P_{\rm max} = 97\%$.}
\end{figure*} 

\clearpage
\begin{figure*}
   \centering
   \includegraphics[width=12cm]{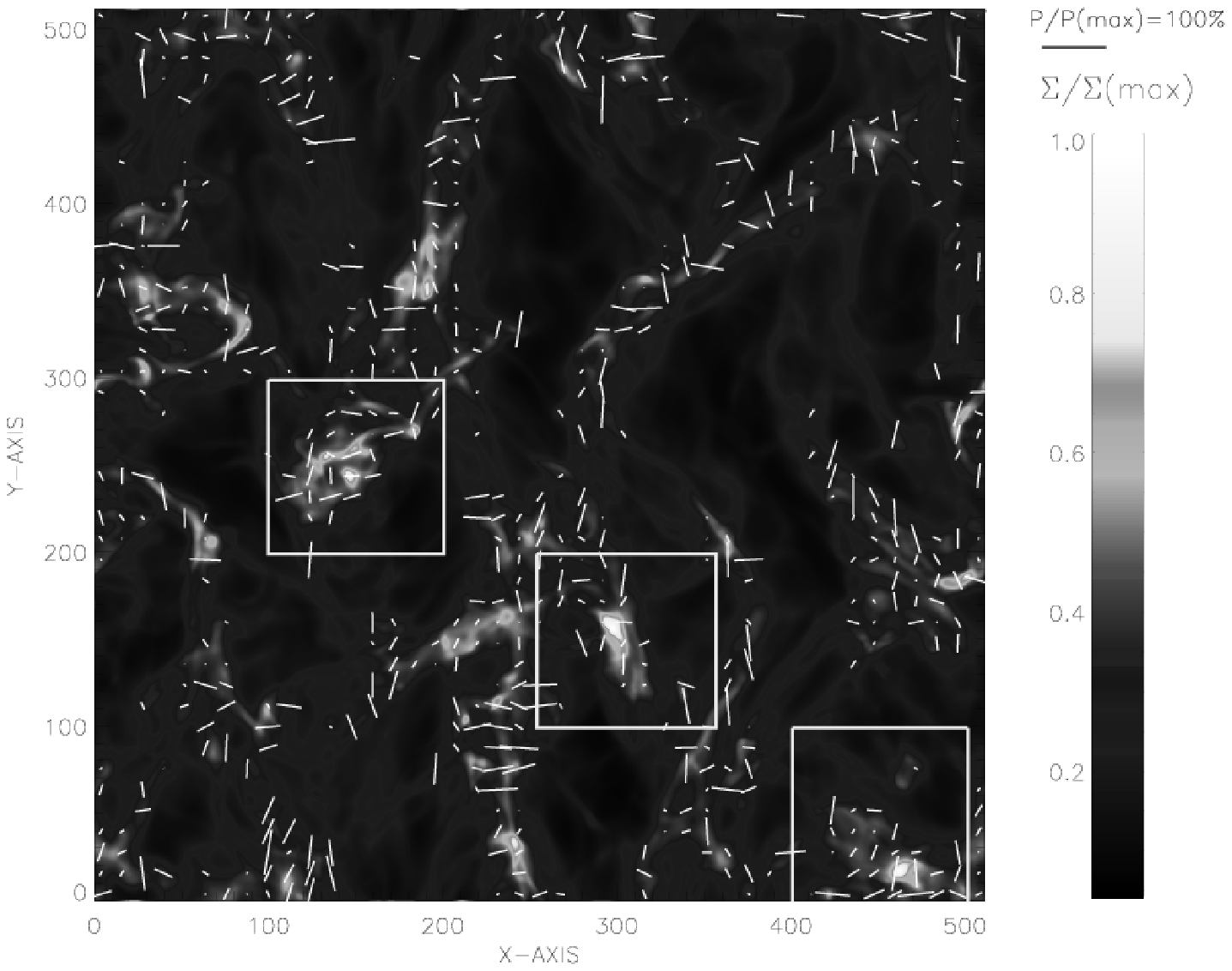}
   \caption{Polarization of emission and column density maps for Model 3 ($M_{\rm S} \sim 7.0$  and $M_{\rm A} \sim 0.7$) with ${\bf B}_{ext}$ parallel to the line of sight. The complete map (512x512 pix) ({\it Upper-left}) and the zoomed regions (100x100 pix). The sensitivity in simulated observations is assumed to be 0.3 of the maximum emission. Here, regions where the signal is less than 0.3 do not show polarization vectors, and $P_{\rm max} = 85\%$.}
\end{figure*} 

\clearpage
\begin{figure*}
   \centering
   \includegraphics[width=12cm]{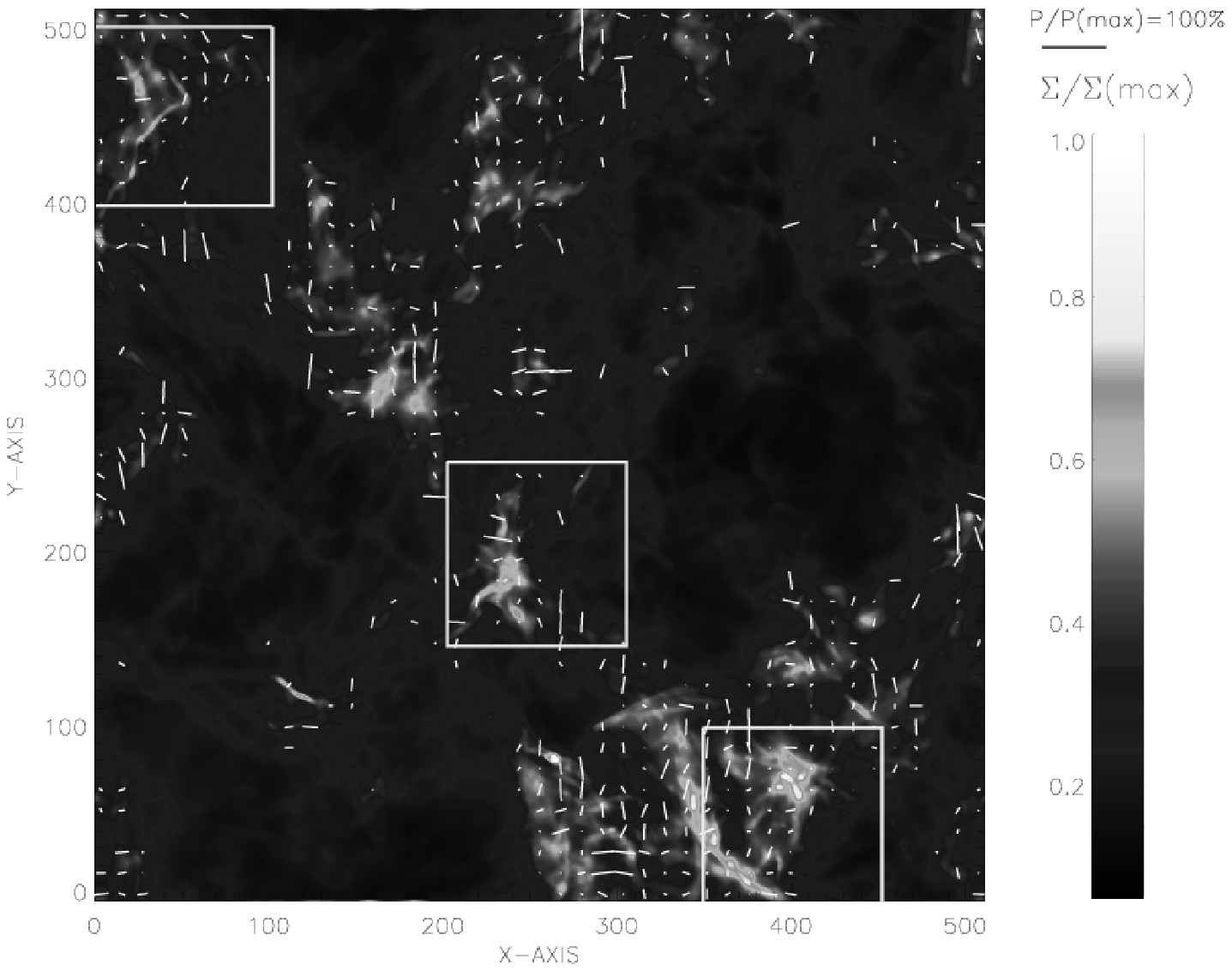}
   \caption{Polarization of emission and column density maps for Model 4 ($M_{\rm S} \sim 7.0$  and $M_{\rm A} \sim 2.0$) with ${\bf B}_{ext}$ 
perpendicular to the line of sight. The complete map (512x512 pix) ({\it Upper-left}) and the zoomed regions (100x100 pix). The sensitivity in simulated observations is assumed to be 0.3 of the maximum emission. Regions where the signal is less than 0.3 do not show polarization vectors. Here, regions where the signal is less than 0.3 do not show polarization vectors, and $P_{\rm max} = 76\%$.}
\end{figure*}

\clearpage
\begin{figure}
   \centering
   \includegraphics[width=13cm]{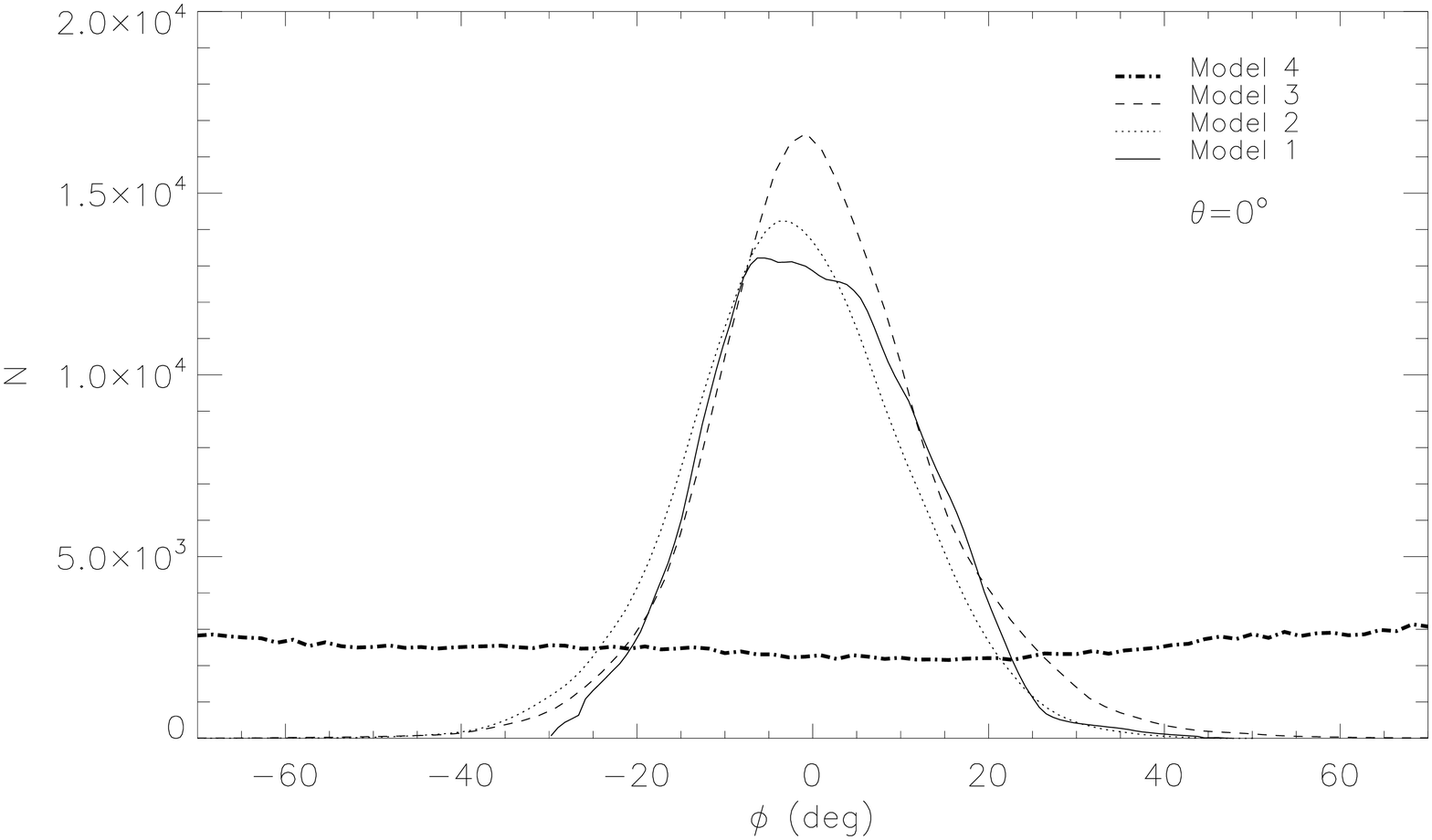}\\[10pt]
   \includegraphics[width=13cm]{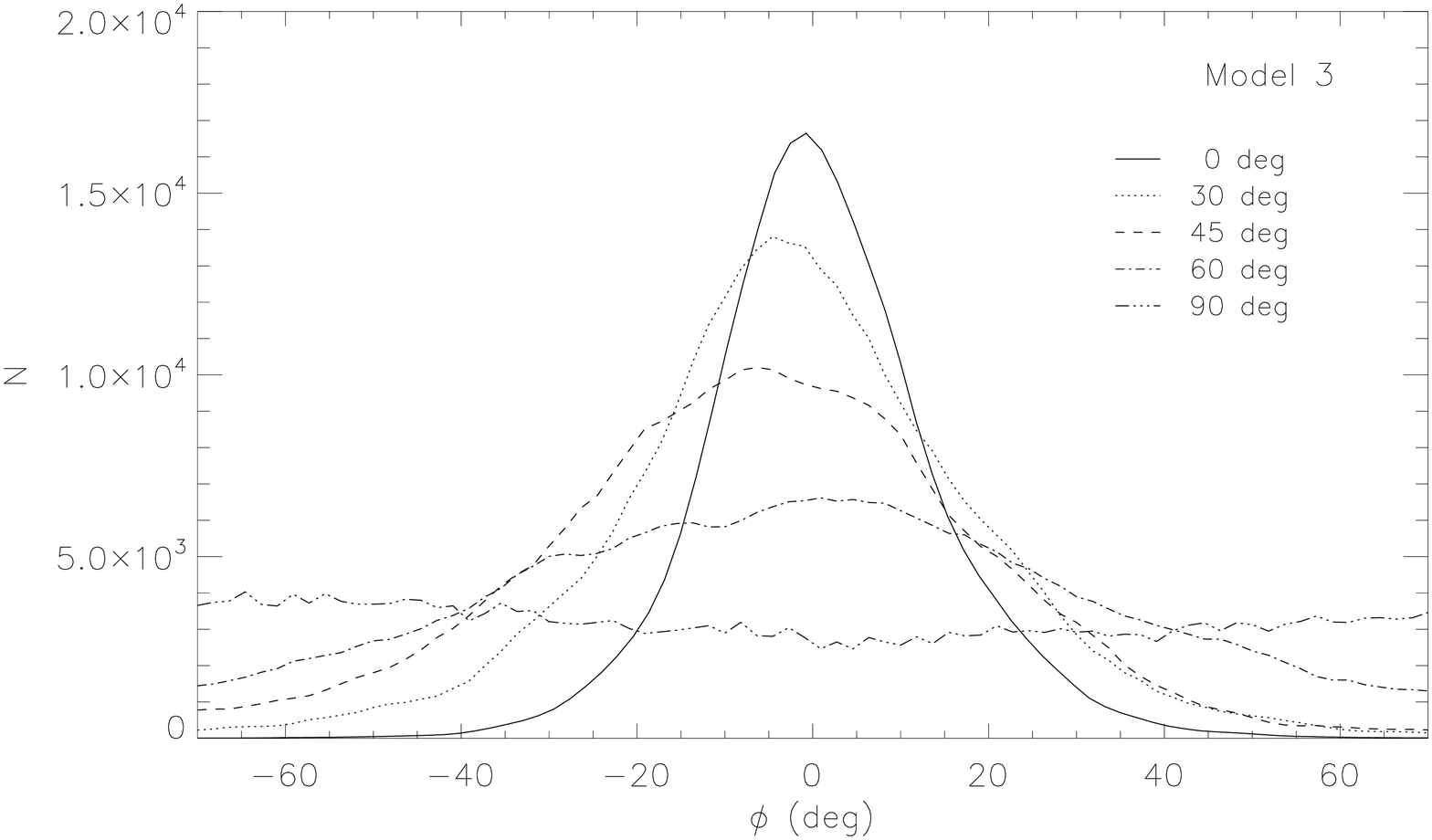}
   \caption{Histograms of polarization angle of the different models with $\theta=0$ ({\it up}), and for Model 3 and different magnetic field orientations in respect to the line of sight (angles $\theta$) ({\it bottom}).}
\end{figure}

\clearpage
\begin{figure}
   \centering
   \includegraphics[width=11cm]{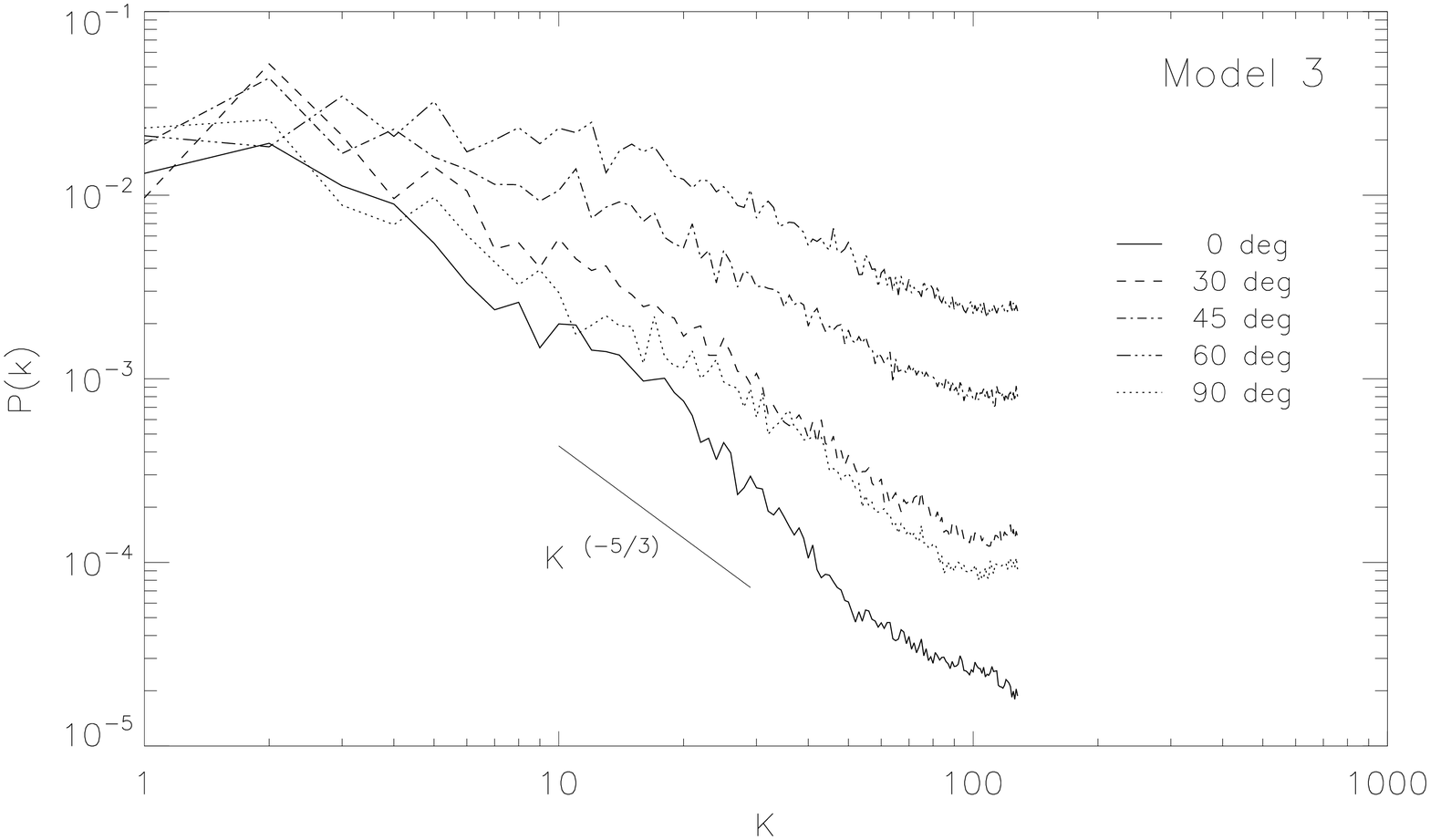}\\[0pt]
   \includegraphics[width=11cm]{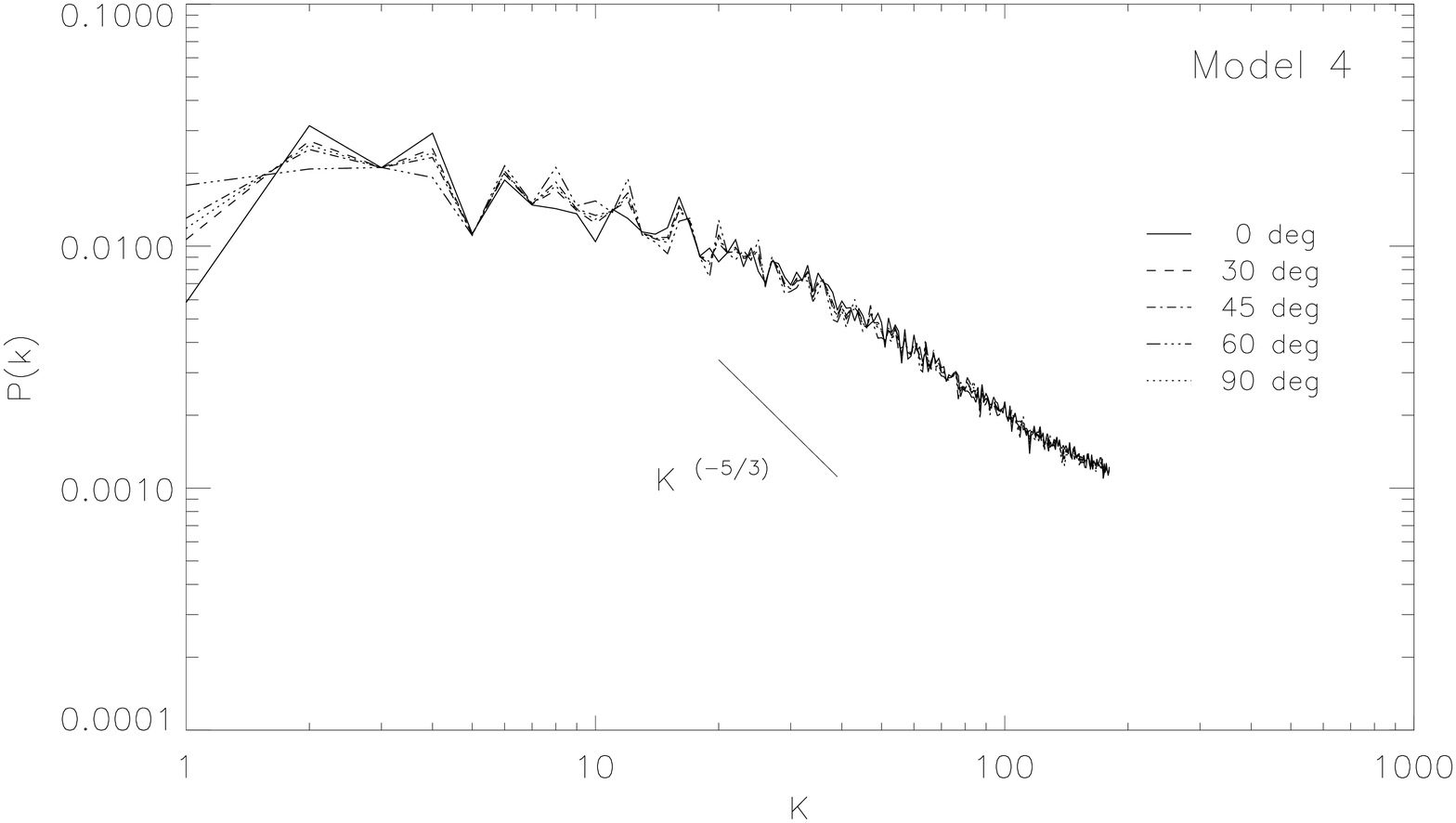}\\[0pt]
   \includegraphics[width=11cm]{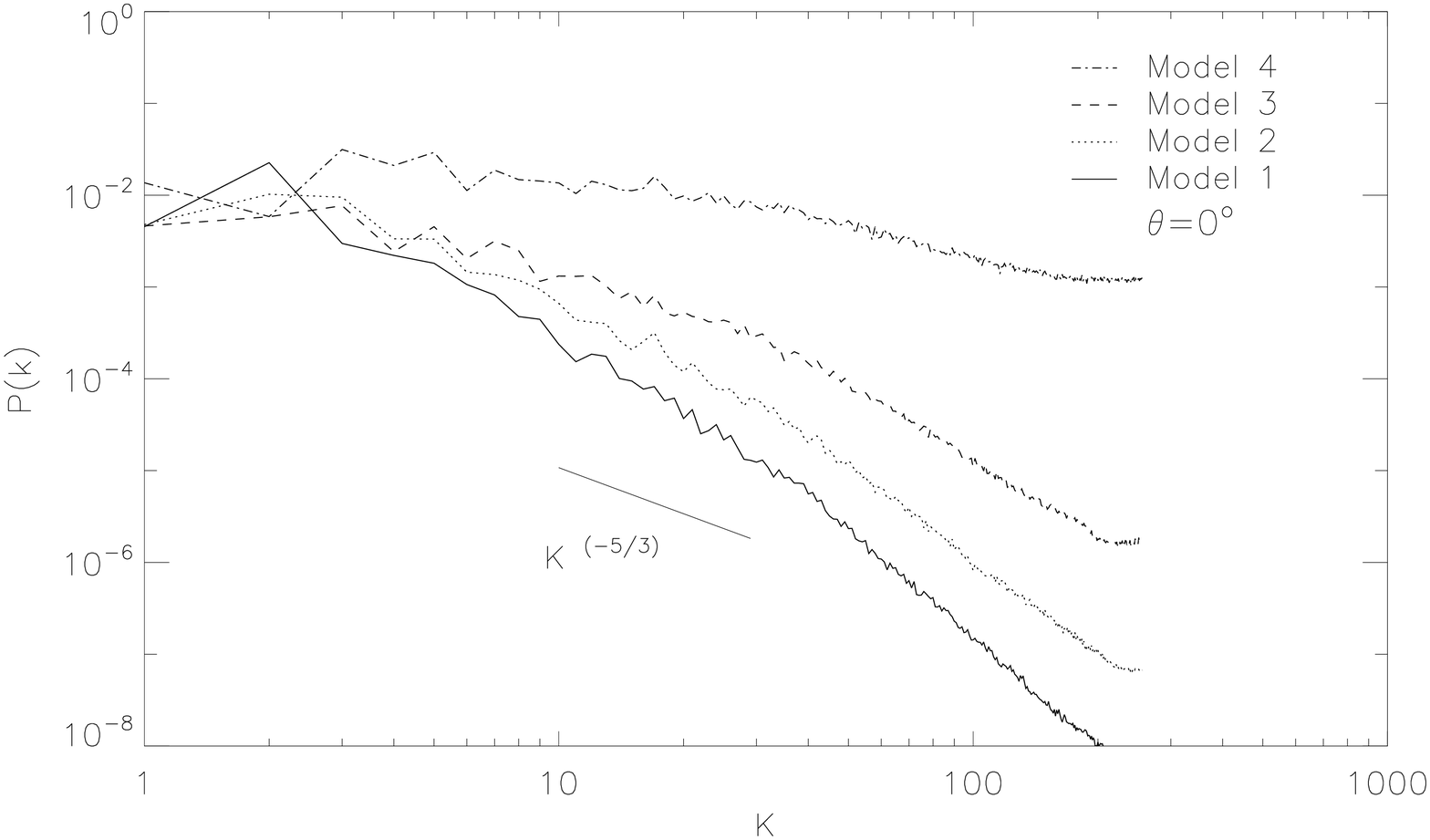}
   \caption{Spectra of polarization angle for Model 3 with different magnetic field orientations regarding the line of sight (angles $\theta$) ({\it up}), Model 4 with different $\theta$  ({\it middle}) and for the different models with magnetic field perpendicular to the line of sight ($\theta = 0$) ({\it bottom}).}
\end{figure}

\clearpage
\begin{figure}
   \centering
   \includegraphics[width=10cm]{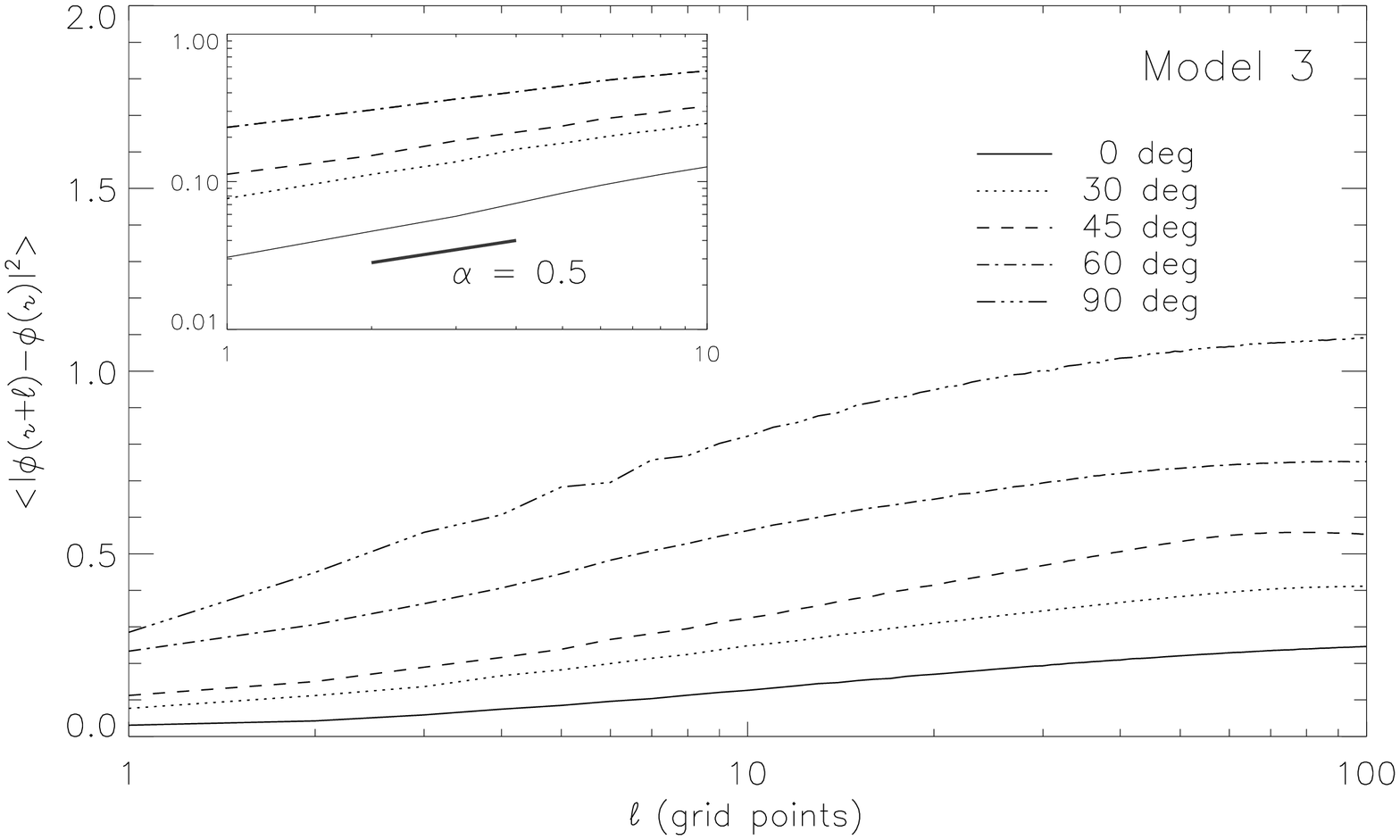}\\[0pt]
   \includegraphics[width=10cm]{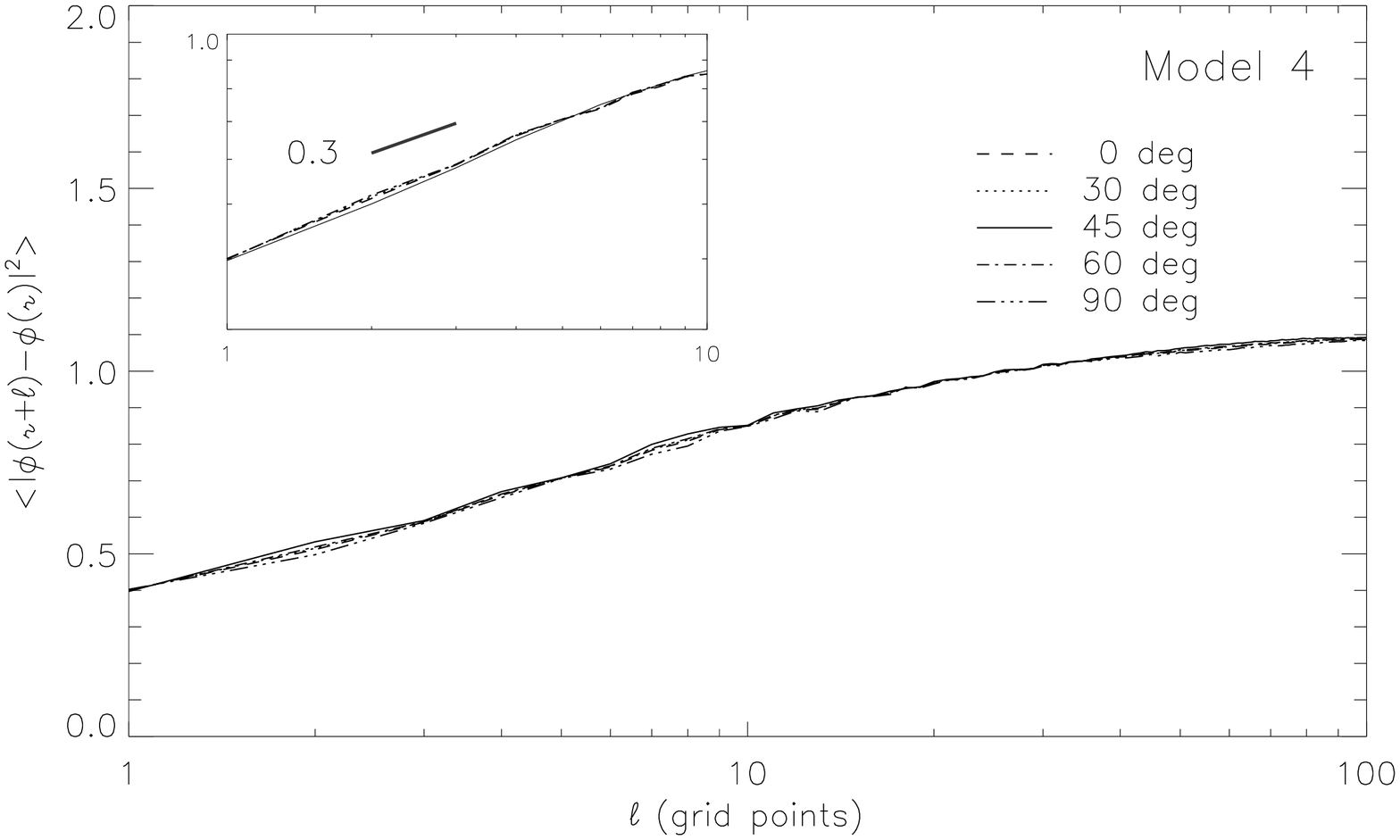}\\[0pt]
   \includegraphics[width=10cm]{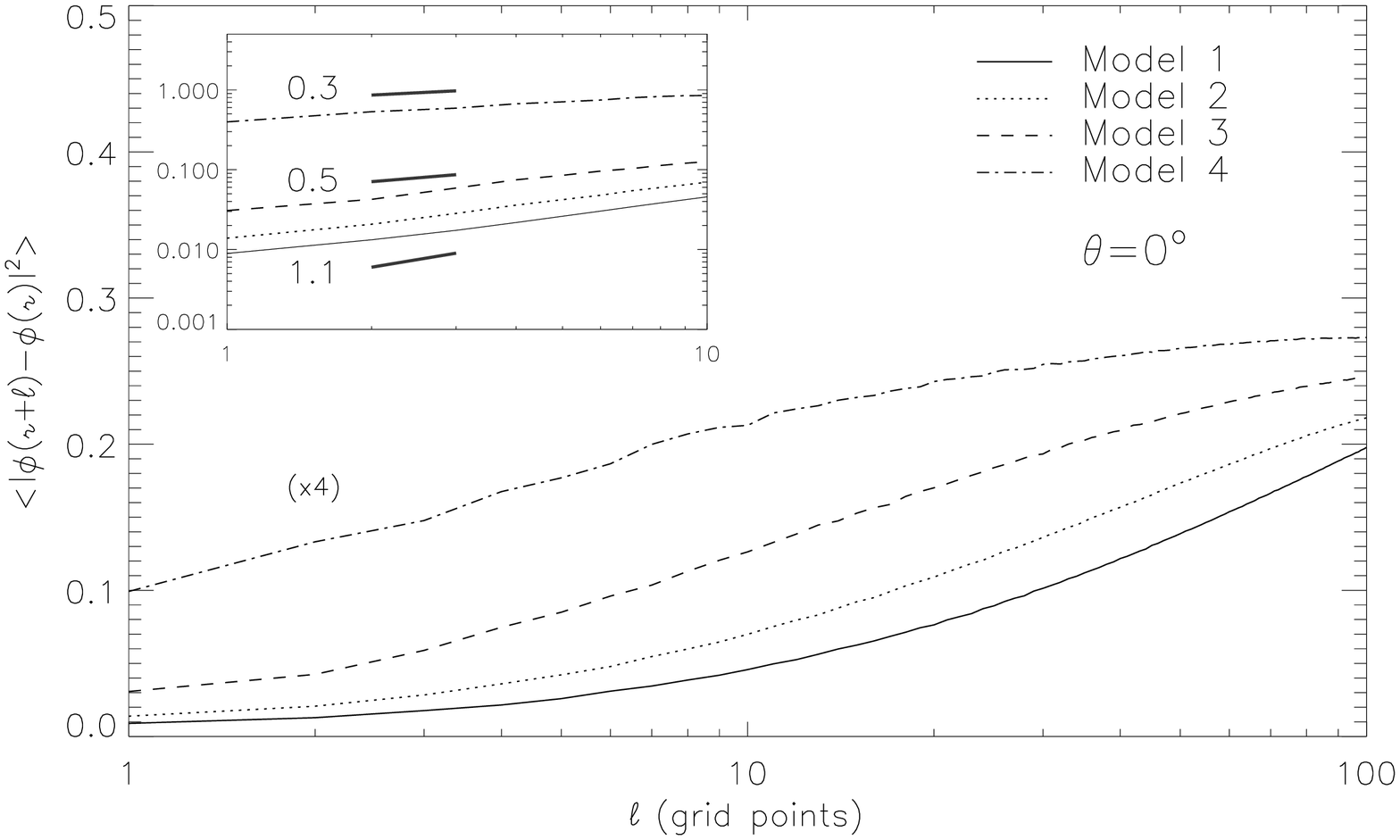}
   \caption{Structure functions of polarization angle for Model 3 with different magnetic field orientations 
regarding the line of sight (angles $\theta$) ({\it up}), Model 4 with different $\theta$  ({\it middle}) and for the different models with magnetic field perpendicular to the line of sight ($\theta = 0$) ({\it bottom}).}
\end{figure}

\clearpage
\begin{figure}
   \centering
   \includegraphics[width=12.0cm]{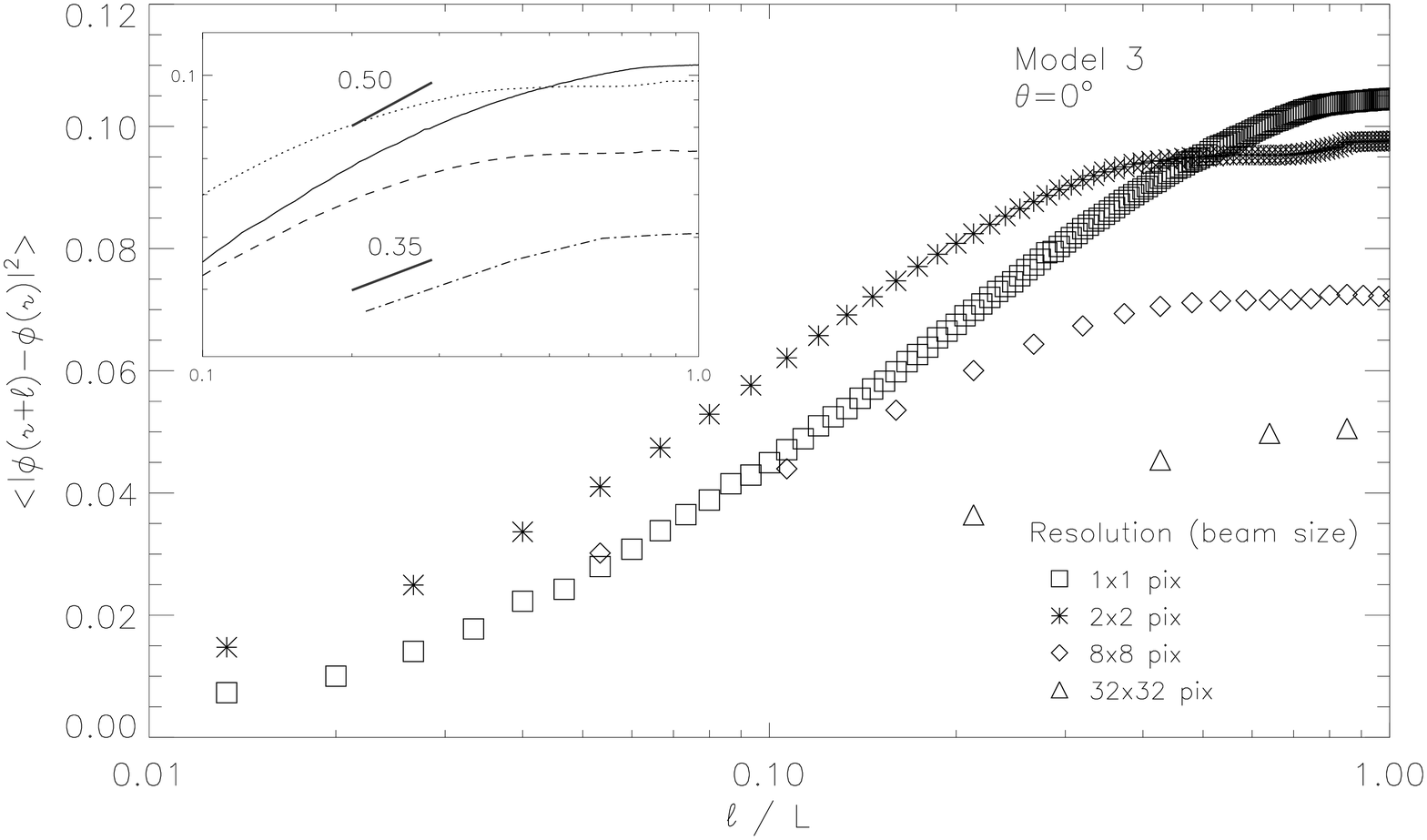}\\[20pt]
   \includegraphics[width=12.0cm]{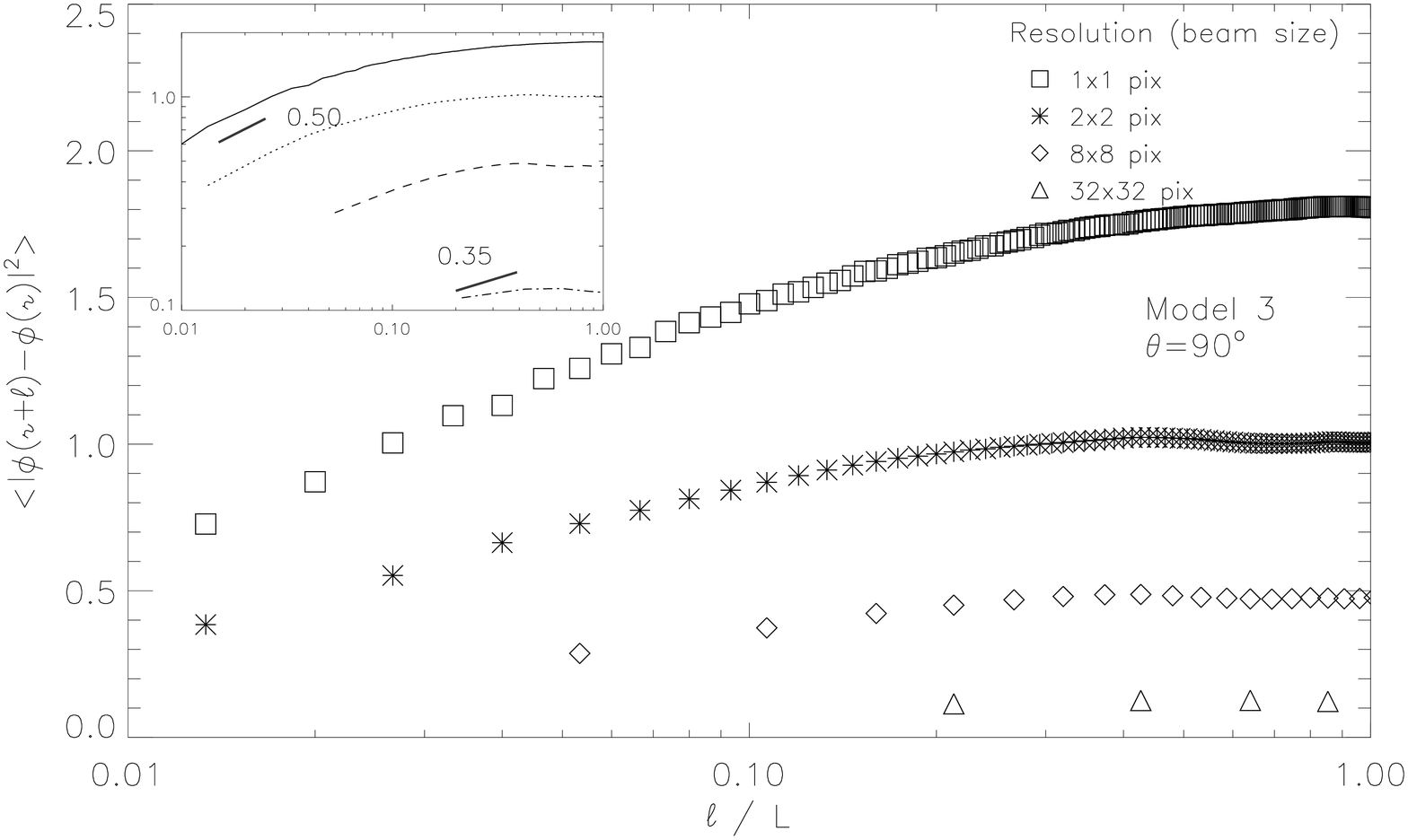}
   \caption{Structure functions of polarization angles for Model 3 considering $B_{\rm ext}$ perpendicular ({\it up}) and parallel to the line of sight ({\it bottom}).}
\end{figure}

\clearpage
\begin{figure}
   \centering
   \includegraphics[width=11cm]{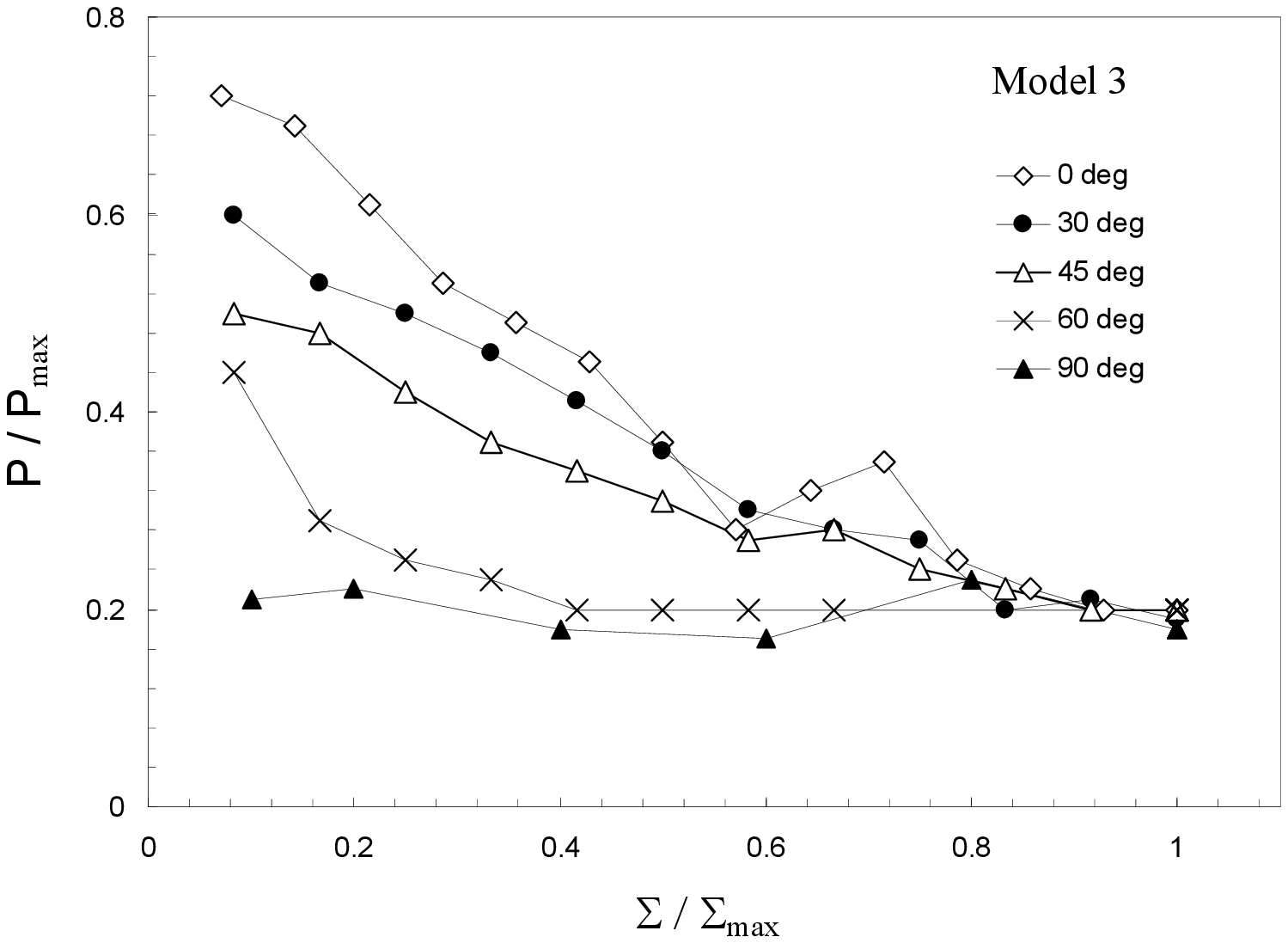}\\[20pt]
   \includegraphics[width=11cm]{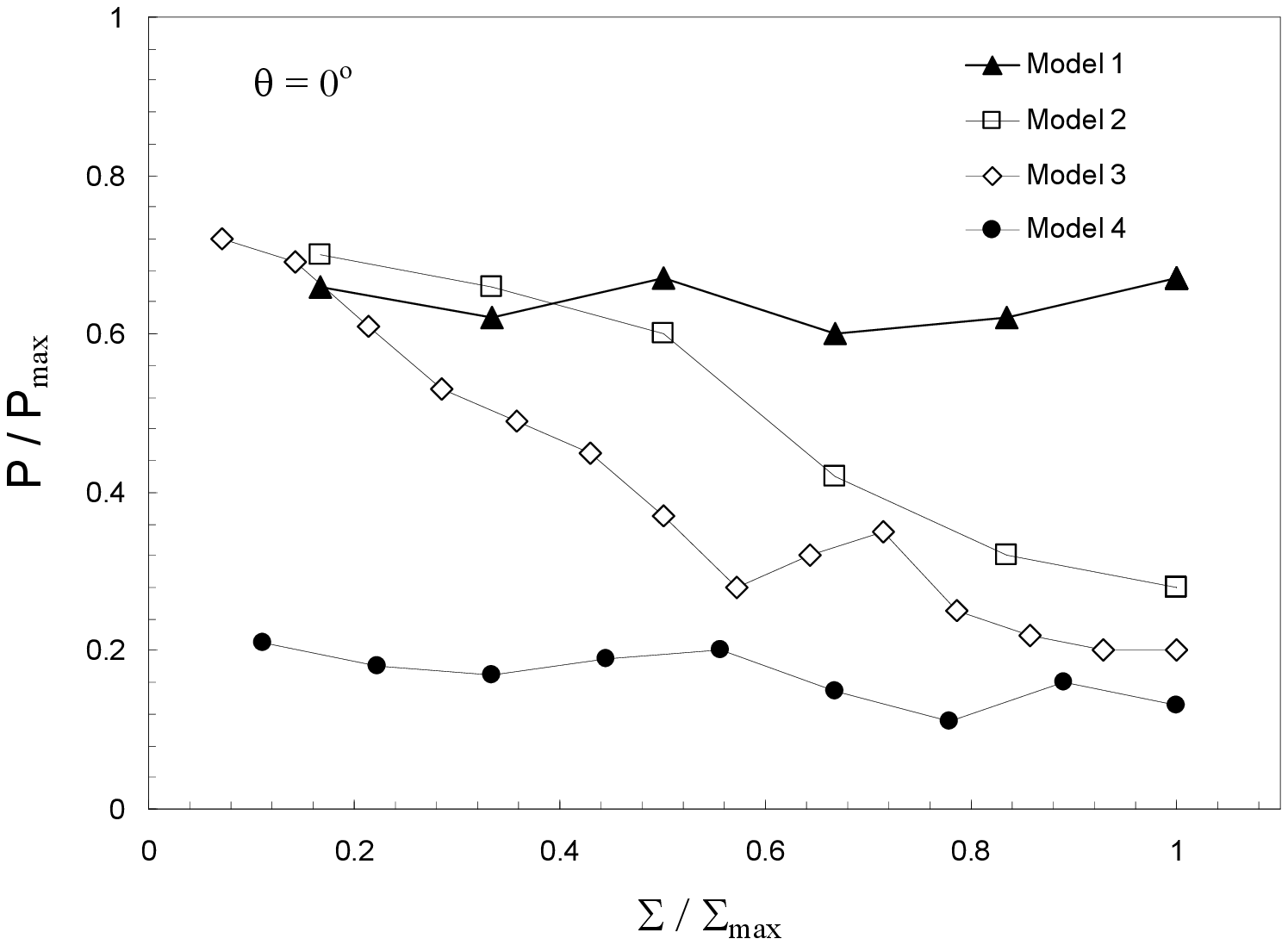}
   \caption{Correlation between averaged polarization degree and the column density for Model 3 with different magnetic field orientations regarding the LOS ({\it up}), and for the different models with $\theta = 0$ ({\it bottom}). $P_{\rm max}$ is 100\%, 98\%, 97\% and 85\% for Model 1, 2, 3 and 4, respectively.}
\end{figure}

\clearpage
\begin{figure}
   \centering
   \includegraphics[width=10cm]{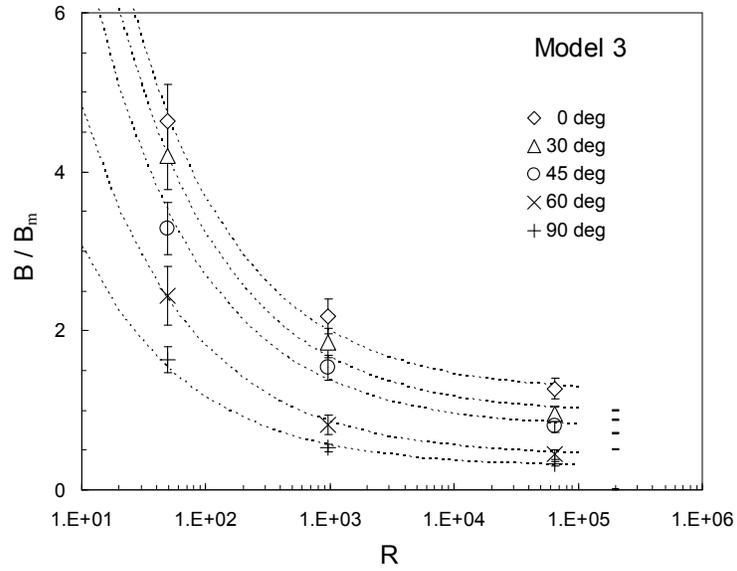}
   \caption{CF method calculation for Model 3 with different inclinations with respect to the line of sight. The dotted 
lines represent the fittings using Eq.\ (10). The traces indicate the expected value $B^{\rm m} \cos \theta$.}
\end{figure}

\clearpage
\begin{figure}
   \centering
   \includegraphics[width=10cm]{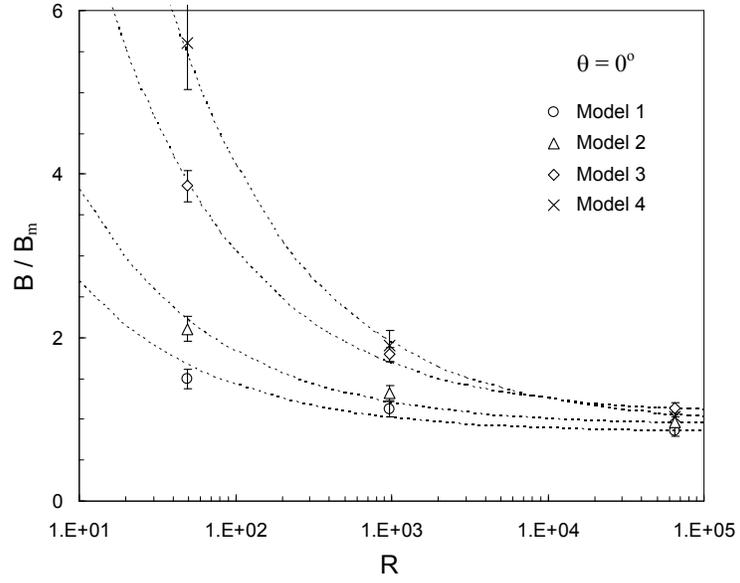}
   \caption{CF method calculation for the different models with $\theta=0$. The dotted 
lines represent the fittings using Eq.\ (10).}
\end{figure}

\clearpage
\begin{figure}
   \centering
   \includegraphics[width=10.0cm]{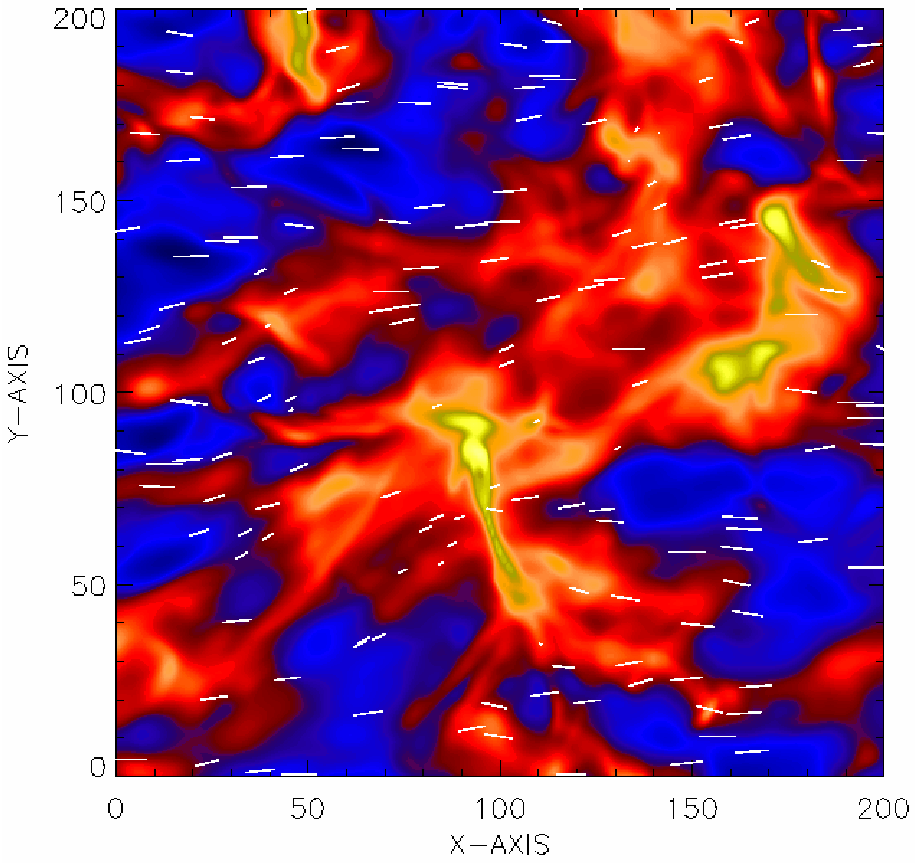}\\[0pt]
   \includegraphics[width=12.0cm]{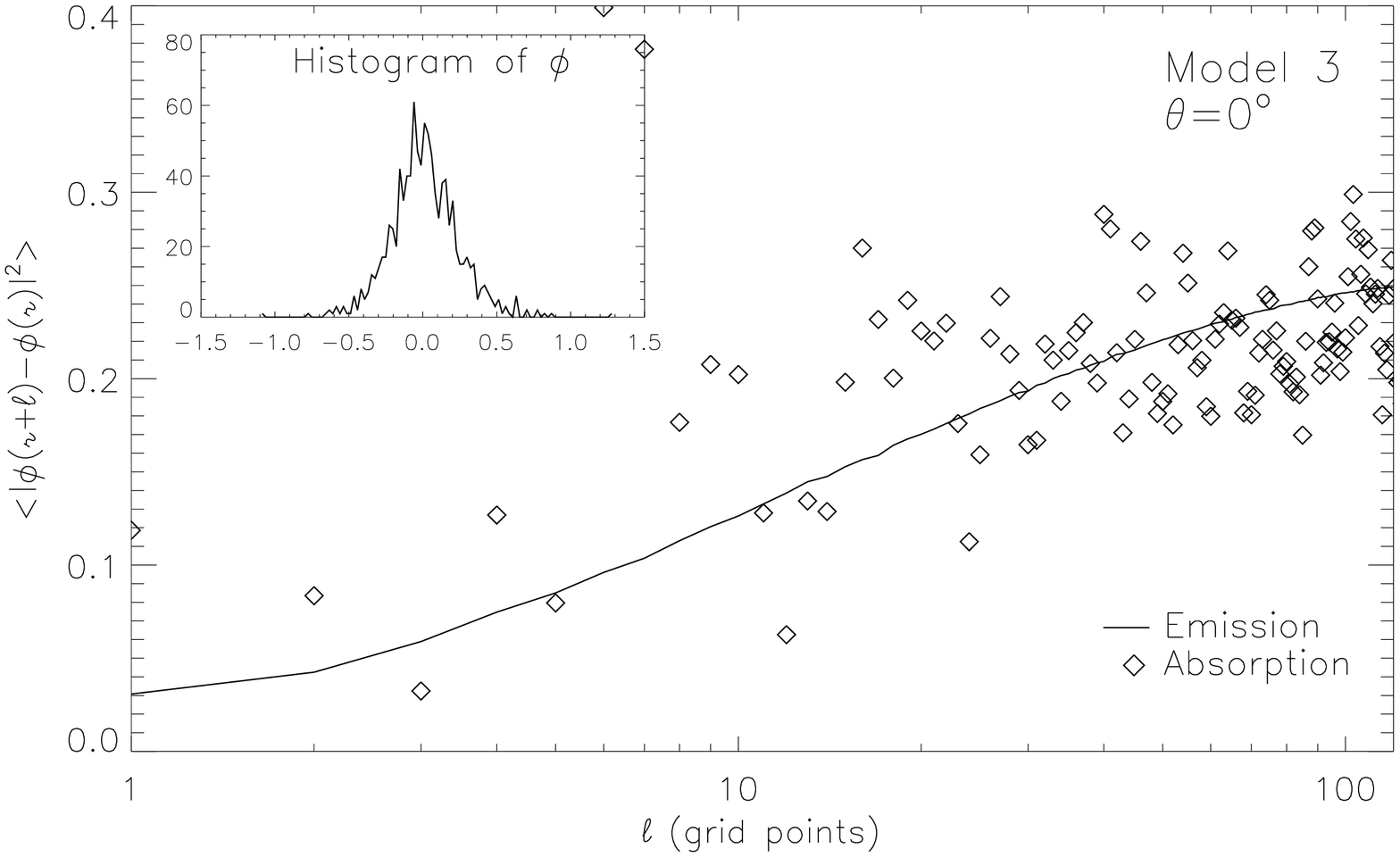}
   \caption{Polarization map of absorbed radiation from 1000 randomly positioned background stars. {\it Up:} column density and polarization vectors of a zoomed region of 200 x 200 pixels for Model 3 with $\theta = 0\degr$. {\it Bottom}: histogram of polarization angle for Model 3 with $\theta = 0\degr$, and its structure function (squares) compared to the emission SF (solid line).}
\end{figure}


\begin{thebibliography}{}

%\bibitem[Bethell et al.(2007)]{bethell07} Bethell, T. J., Chepurnov, A., Lazarian, A. \& Kim, J. 2007, \apj, 663, 1055
\bibitem[Arce et al.(1998)]{arce98} Arce, H. G., Goodman, A. A., Bastien, P., Manset, N. \& Sumner, M. 1998, \apj, 499, 93
%\bibitem[Cho, Lazarian \& Vishniac(2002)]{cho02} Cho, J., Lazarian, A. \& Vishniac, E. T. 2002, \apj, 564, 291
\bibitem[Cortes, Crutcher \& Watson(2005)]{cortes05} Cortes, P. C., Crutcher, R. M. \& Watson, W. 2005, \apj, 628, 780
\bibitem[Cho \& Lazarian(2005)]{cho05} Cho, J. \& Lazarian, A. 2005, \apj, 631, 361
\bibitem[Chandrasekhar \& Fermi(1953)]{chandra53} Chandrasekhar, S. \& Fermi, E. 1953, \apj, 118, 113
\bibitem[Crutcher(1999)]{cru99} Crutcher, R. M. 1999, \apj, 520, 706
\bibitem[Crutcher et al.(1999)]{cru99b} Crutcher, R. M., Roberts, D. A., Troland, T. H. \& Goss, W. M. 1999, \apj, 515, 275
\bibitem[Dolginov \& Mytrophanov(1996)]{dolg76} Dolginov, A. Z. \& Mytrophanov, I. G. 1976, \apss, 43, 257
\bibitem[Dotson(1996)]{dotson96} Dotson, J. 1996, \apj, 470, 566
\bibitem[Draine \& Weingartner(1996)]{draine96} Draine, B. T. \& Weingartner, J. C. 1996, \apj, 470, 551
\bibitem[Draine \& Weingartner(1997)]{draine97} Draine, B. T. \& Weingartner, J. C. 1997, \apj, 480, 633
\bibitem[Esquivel \& Lazarian(2005)]{esquivel05} Esquivel, A. \& Lazarian, A. 2005, \apj, 631, 320
\bibitem[Garwood \& Jones(1987)]{garwood87} Garwood, R. \& Jones, T. J. 1987, \pasp, 99, 453
\bibitem[Girart, Crutcher \& Rao(1999)]{girart99} Girart, J. M., Crutcher, R. M. \& Rao, R. 1999, \apj, 525, 109
\bibitem[Girart et al.(2004)]{girart04} Girart, J., Greaves, J. M., Crutcher, R. M. \& Lai, S.-P. 2004, \apss, 292, 119
\bibitem[Girart, Rao \& Marrone(2006)]{girart06} Girart, J. M., Rao, R. \& Marrone, D. P. 2006, Science, 313, 812
\bibitem[Goldreich \& Kylafis(1981)]{gold81} Goldreich, P. \& Kylafis, N. D. 1981, \apj, 243, 75
\bibitem[Goldreich \& Kylafis(1982)]{gold82} Goldreich, P. \& Kylafis, N. D. 1982, \apj, 253, 606
\bibitem[Gon\c calves, Galli \& Walmsley(2005)]{goncalves05} Gon\c calves, J., Galli, D. \& Walmsley, M. 2005, \aap, 430, 979
\bibitem[Greaves, Holland \& Dent(2002)]{greaves02} Greaves, J. S., Holland, W. S. \& Dent, W. R. F. 2002, \apj, 578, 224
\bibitem[Heiles \& Troland(2005)]{heiles05} Heiles, C. \& Troland, T. H. 2005, \apj, 624, 773
\bibitem[Heitsch et al.(2001)]{heitsch01} Heitsch, F., Zweibel, E., Mac Low, M. M., Li, P. S., \& Norman, M. L. 2001, \apj, 561, 800
\bibitem[Hildebrand et al.(2000)]{hil00} Hildebrand, R. H., Davidson, J. A., Dotson, J. L., Dowell, C. D., Novak, G., \& Vaillancourt,
J. E. 2000, \pasp, 112, 1215
\bibitem[Hildebrand et al.(1999)]{hil99} Hildebrand, R. H., Dotson, J. L., Dowell, C. D., Schleuning, D. A. \& Vaillancourt, J. E. 1999, \apj, 516, 834
\bibitem[Hoang \& Lazarian(2008)]{hoang08} Hoang, T. \& Lazarian, A. 2008, \mnras, in press
\bibitem[Kritsuk et al.(2007)]{kritsuk07} Kritsuk, A. G., Norman, M. L., Padoan, P. \& Wagner, R. 2007, \apj, 665, 416
\bibitem[Kowal, Lazarian \& Beresniak(2007)]{kowal07} Kowal, G., Lazarian, A. \& Beresniak, A. 2007, \apj, 658, 423
\bibitem[Lai, Girart \& Crutcher(2003)]{lai03} Lai, S.-P., Girart, J. M. \& Crutcher, R. M. 2003, \apj, 598, 392
\bibitem[Lazarian, Goodman \& Myers(1997)]{laz97} Lazarian, A., Goodman, A. A. \& Myers, P. C. 1997, \apj, 490, 273
\bibitem[Lazarian \& Pogosyan(2000)]{lp00} Lazarian, A. \& Pogosyan, D. 2000, \apj, 537, 720
\bibitem[Lazarian \& Cho(2004)]{lazcho04} Lazarian, A. \& Cho, J. 2004, \apss, 292, 29
\bibitem[Lazarian, Vishniac \& Cho(2004)]{laz04} Lazarian, A., Vishniac, E. T. \& Cho, J. 2004, \apj, 603, 180
\bibitem[Lazarian(2007)]{laz07} Lazarian, A. 2007, \jqsrt, 106, 225
\bibitem[Lazarian \& Hoang(2007a)]{lazho07a} Lazarian, A. \& Hoang, T. 2007a, \mnras, 378, 910
\bibitem[Lazarian \& Hoang(2007b)]{lazho07b} Lazarian, A. \& Hoang, T. 2007b, \apj, 669, 77
\bibitem[Matthews \& Wilson(2002)]{matthews02} Matthews, B. C. \& Wilson, C. D. 2002, \apj, 574, 822
\bibitem[Myers \& Goodman(1991)]{myers91} Myers, P. C. \& Goodman, A. A. 1991, \apj, 373, 509
\bibitem[Ostriker, Stone \& Gammie(2001)]{ostriker01} Ostriker, E. C., Stone, J. M. \& Gammie, C. F. 2001, \apj, 546, 980
\bibitem[Padoan et al.(2001)]{padoan01} Padoan, P., Goodman, A., Draine, B. T., Juvela, M., Nordlund, A. \& Rognvaldsson, o. E. 2001, \apj, 559, 1005
\bibitem[Padoan \& Norlund(2002)]{padoan02} Padoan, P. \& Norlund, A. 2002, \apj, 576, 870
\bibitem[Pelkonen, Juvela \& Padoan(2007)]{pelkonen07} Pelkonen, V.-M., Juvela, M. \& Padoan, P. 2007, \aap, 461, 551
\bibitem[Schleuning(1998)]{sch98} Schleuning, D. A. 1998, \apj, 493, 811
\bibitem[Shiode et al.(2006)]{shiode06} Shiode, Joshua H., Clemens, D. P., Janes, K. A., Pinnick, A. \& Taylor, B. 2006, \baas, 209, 156
\bibitem[Vall\'ee \& Fiege(2007)]{vallee07} Vall\'ee, J. P. \& Fiege, J. D. 2007, \aj, 134, 628
\bibitem[Whittet et al.(2008)]{whittet08} Whittet, D. C. B., Hough, J. H., Lazarian, A. \& Hoang, T. 2008, \apj, in press
\bibitem[Wolf, Launhardt \& Henning(2003)]{wolf03} Wolf, S., Launhardt, R. \& Henning, T. 2003, \apj, 592, 233
\bibitem[Yan \& Lazarian(2006)]{yan06} Yan, H. \& Lazarian, A. 2006, \apj, 653, 1292
\bibitem[Yan \& Lazarian(2007a)]{yan07a} Yan, H. \& Lazarian, A. 2007a, \apj, 657, 618
\bibitem[Yan \& Lazarian(2007b)]{yan07b} Yan, H. \& Lazarian, A. 2007b, astro-ph:0711.0926
\bibitem[Zweibel(1990)]{zweibel90} Zweibel, E. G. 1990, \apj, 362, 545

\end{thebibliography}
\end{document}